\documentclass[preprint,showpacs,preprintnumbers,amsmath,amssymb]{revtex4}
\usepackage{amssymb}
\usepackage{graphicx}
\usepackage{subfigure}
\usepackage{dcolumn}
\usepackage{bm}
\usepackage{amsmath}

\newcommand{\bra}[1]{\left\langle #1 \right|}
\newcommand{\ket}[1]{\left| #1 \right\rangle}
\newcommand{\braket}[2]{\left\langle #1 | #2 \right\rangle}

\newcommand{\cre}[2]{\hat{#1}^{\dag}_{#2}}
\newcommand{\ann}[2]{\hat{#1}_{#2}}
\renewcommand{\vec}[1]{\mathbf{#1}}
\newcommand{\abs}[1]{\left|#1\right|}
\begin{document}

\title{Dressed state approach to matter wave mixing of bosons}

\author{E. Rowen}
\author{R. Ozeri}

\author{N. Katz}
\author{R. Pugatch}
\author{N. Davidson}

\affiliation{Department of Physics of Complex Systems,\\
Weizmann Institute of Science, Rehovot 76100, Israel}
\date{\today}

\begin{abstract}

A dressed state approach to mixing of bosonic matter waves is
presented. Two cases are studied using this formalism. In the
first, two macroscopically populated modes of atoms (two-wave
mixing) are coupled through the presence of light. In the second
case, three modes of Bogoliubov quasiparticles (three-wave mixing)
are coupled through s-wave interaction. In both cases wave mixing
induces oscillations in the  population of the different modes
that decay due to interactions. Analytic expressions for the
dressed basis spectrum and the evolution of the mode populations
in time are derived both for resonant mixing and non-resonant
mixing. Oscillations in the population of a given mode are shown
to lead to a splitting in the decay spectrum of that mode, in
analogy to the optical Autler-Townes splitting in the decay
spectrum of a strongly driven atom. These effects cannot be
described by a mean-field approximation.
\end{abstract}


 \maketitle

\section{Introduction}

Wave mixing is a well known phenomenon which occurs in nonlinear
systems. The realization of Bose-Einstein condensates (BEC) in
atomic vapor paved the path to the experimental study of mixing of
matter waves
\cite{phillips-4wm,ketterle-coherent_col,ketterle-4wm}, in which
the s-wave interactions are the cause for the required
nonlinearity. This exciting new effect is  analogous to the wave
mixing of optical modes in nonlinear processes such as parametric
down-conversion \cite{downconversion}. As in the optical case,
atomic wave mixing is predicted to generate number squeezed
states. These states are usually referred to as quantum states
with a well defined number and a spread in the phase,  contrary to
the BEC itself, which is described well by a classical wave in the
sense that both number and phase are well defined. These squeezed
states can be used to perform sub-shotnoise measurements in
interferometry experiments \cite{kasevich-subshotnoise}.

In previous work, the nonlinear mixing of Bogoliubov
quasiparticles was studied using the Gross-Pitaevskii equation
(GPE) \cite{burnett-nonlin}. In the high momentum limit, in which
most experiments are performed, atomic four-wave mixing was
studied using the GPE \cite{band-4wm}, and  analyzed directly in
the quantum many body formalism using angular momentum for mixing
of different hyperfine modes \cite{meystre-4wm}. The coherence in
wave mixing of two light modes with two atomic modes was studied
experimentally \cite{ketterle-coherent_amp} and demonstrated when
 only one atomic mode and one light mode are initially
populated \cite{ketterle-SR,ketterle-BackwardSR}.

Many dynamical effects in ultra-cold atoms can be described by the
time dependent GPE, which  governs the evolution of a
macroscopically occupied wavefunction. This approach cannot be
used in order to study quantum effects where number and phase
cannot both be determined at once. Here we develop a dressed state
model \cite{ours-3wm}, which unlike the GPE, does \emph{not}
assume all atoms are in the same single particle state. We use
this model to describe effects that are beyond mean-field, such as
dephasing due to quantum uncertainty in Rabi oscillations between
momentum modes, and the decay of a matter wave due to interaction
with the quasi-continuum of unoccupied modes.

Weak interactions are usually taken into account by the Bogoliubov
transformation from the atomic basis to a basis of quasiparticles.
Interactions between quasiparticles and their decay were first
analyzed by Beliaev \cite{beliaev}, and experimentally verified
\cite{foot-beliaev,ours-beliaev}. In this paper we study wave
mixing between macroscopically occupied quasiparticle modes. For
simplicity we
 consider throughout the paper only homogeneous condensates at $T=0$.
 We find that interactions between quasiparticles
can lead to new phenomena in matter waves, analogous to phenomena
studied in connection to interactions of light waves with matter,
such as a splitting in the decay spectrum, power broadening and
energy shifts \cite{API}. We first apply the dressed state
formalism to the simple case of two matter waves coupled by laser
fields. We then adapt it to solve the more interesting case of
three-wave mixing as a locally perturbed two-wave mixing.

Treating the wave mixing problem in the Bogoliubov basis allows us
to consider mixing between quasiparticles in the low momentum
regime, where the basis of atomic modes (used e.g. to describe
four-wave mixing of matter waves \cite{phillips-4wm,ketterle-4wm})
is not appropriate  due to the large $-\vec{k}$ atomic component
of a quasiparticle with momentum $\vec{k}$. Such a low momentum
regime is of particular interest since both the density of states
and quantum interference effects yield very low decay rates
\cite{ours-beliaev,enhance-supp}, and is therefore more suitable
experimentally for the study of relative number squeezing and
entanglement between matter waves. In the high momentum regime,
three-wave mixing coincides with the conventional four-wave mixing
of matter waves as long as the condensate, which is the fourth
wave, is not substantially depleted.

The layout of this paper is as follows: First, in section
\ref{sec:hamiltonians} we review the Bogoliubov transformation and
 the wave mixing Hamiltonians for two physical
scenarios. We describe both the mixing of Bogoliubov
quasiparticles due to the s-wave interaction, and the mixing of
atomic modes which are coupled by laser fields, and in which case
interactions may be included as a perturbation. We also derive the
perturbative dynamics and energy shifts due to the interactions
between Bogoliubov quasiparticles. Next, in section \ref{sec:2wm},
we solve the two-wave mixing Hamiltonian. For noninteracting
atoms, this is in fact a single particle problem. However, we use
the second-quantized formalism, which will become a necessity once
interactions are introduced, and nonlinear dynamics emerge. We
compare this formalism to the two-mode Gross Pitaevskii model
\cite{ami}, and discuss the dynamical instability, as well as a
decay due to quantum uncertainty in the number of quasiparticles.
In section \ref{sec:3wm} we turn to the nonlinear three-wave
mixing between Bogoliubov quasiparticles, using the framework
introduced in section \ref{sec:2wm}. We  study the dynamics and
spectrum of resonant and non-resonant three-wave mixing. We find
an analytic approximation which is in good agreement with
numerical diagonalization, and derive the scaling laws which let
us describe a realistic system by matrices which can be
diagonalized on a personal computer. Finally, in section
\ref{sec:damping}, we discuss the  decay of a quasiparticle
coupled both to a quasi-continuum and an additional
macroscopically occupied quasiparticle mode. We find a splitting
in the decay spectrum in analogy to the Autler-Townes splitting of
the spectrum of an atom undergoing Rabi oscillations. We conclude
with a brief summary and
 examples of other cases to which our formalism can be
applied.

\section{wave mixing Hamiltonians for ultra-cold atoms}\label{sec:hamiltonians}
In this paper we concentrate on a homogeneous system of ultra-cold
bosons, with s-wave interactions. We use this system to derive the
physical Hamiltonians giving rise to two and three-wave mixing.
The three-wave mixing is of Bogoliubov quasiparticles over a
condensate which are inherently coupled by interactions. It
therefore describes interactions correctly even for small momentum
modes, but is not applicable to strong excitations which deplete
the condensate \cite{ours-dephasing}. The two-wave mixing is of
two macroscopically populated momentum modes, which can be
depleted, and are coupled by an external laser field. Interactions
are incorporated as a mean-field shift, thus the two-wave model is
not applicable for small momenta.
\subsection{Wave mixing of Bogoliubov quasiparticles}\label{sec:hamiltonians-3wm}
\begin{figure}[tb]
\begin{center}
\includegraphics[width=8cm]{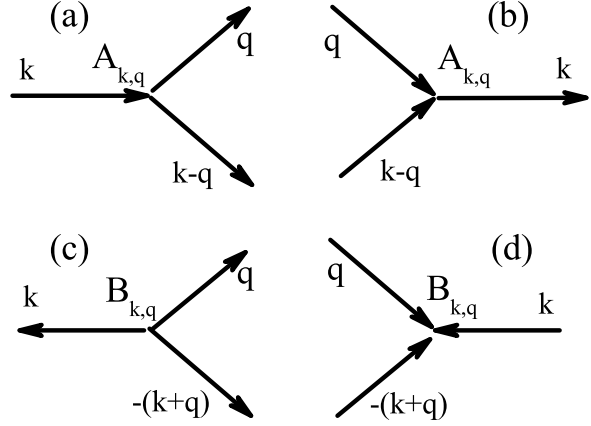}
\end{center}
\caption{The four processes described by $H_{int}$ in Eq.
(\ref{eq:bel_ham}): (a) Beliaev damping (b) Landau damping (c)
spontaneous creation of 3 quasiparticles (d) spontaneous
annihilation of 3 quasiparticles. Only processes (a) and (b) can
be resonant. \label{fig:arrows}}
\end{figure}

 In momentum representation the Hamiltonian governing a system of ultra-cold interacting bosons is given by
 \cite{Fetter-review}:
\begin{equation}\label{eq:H_mb_k_rep}
H=\sum_\vec{k}\frac{\hbar^2
k^2}{2M}\cre{a}{\vec{k}}\ann{a}{\vec{k}}
+\frac{g}{2V}\sum_{\vec{k},\vec{l},\vec{m}}\cre{a}{\vec{k}}\cre{a}{\vec{l}}\ann{a}{\vec{m}}\ann{a}{\vec{k}+\vec{l}-\vec{m}}.
\end{equation}
$g=4\pi\hbar^2 a_s/M$ is a constant proportional to the scattering
length $a_s$,  describing the atom-atom interaction. $V$ is the
volume of the condensate and $M$ is the atomic mass. Assuming a
macroscopic occupation of the ground state $N_0$, thus replacing
both $\ann{a}{0}$ and $\cre{a}{0}$ by $\sqrt{N_0}$ this
Hamiltonian is diagonalized to quadratic order by the Bogoliubov
transformation
\begin{eqnarray}\label{eq:bog_trans}
  \ann{a}{\vec{k}}=u_k\ann{b}{\vec{k}}-v_k\cre{b}{\vec{-k}}\\
\cre{a}{\vec{k}}=u_k\cre{b}{\vec{k}}-v_k\ann{b}{\vec{-k}}.
\end{eqnarray}
When taken to cubic terms in $\ann{b}{}$, Eq.
(\ref{eq:H_mb_k_rep}) can be written as $H=H_0+H_{int}$ with

\begin{equation}\label{eq:bog_ham}
  H_0=E_g+\sum_{\vec{k'}}\epsilon_{k'}\cre{b}{\vec{k'}}\ann{b}{\vec{k'}},
\end{equation}
where $E_g$ is the energy of the Bogoliubov vacuum, and
\begin{equation}\label{eq:bel_ham}
  H_{int}=
  \frac{g\sqrt{N_0}}{2V}\sum_{\vec{k'},\vec{q'}}\left[A_{\vec{k'},\vec{q'}}\left(\ann{b}{\vec{k'}}\cre{b}{\vec{q'}}\cre{b}{\vec{k'-q'}}+
  \cre{b}{\vec{k'}}\ann{b}{\vec{q'}}\ann{b}{\vec{k'-q'}}\right)
  +B_{\vec{k'},\vec{q'}}\left(\cre{b}{\vec{k'}}\cre{b}{\vec{q'}}\cre{b}{-\vec{(k'+q')}}+
  \ann{b}{\vec{k'}}\ann{b}{\vec{q'}}\ann{b}{-\vec{(k'+q')}}\right)\right].
\end{equation}
For weakly excited condensates $H_{int}$ is typically small, and
is usually neglected, as in the derivation of the Bogoliubov
spectrum $\epsilon_k=\sqrt{\frac{\hbar^2 k^2}{2M}(\frac{\hbar^2
k^2}{2M}+ 2\mu)}$, where $\mu=g\rho$ is the chemical potential and
$\rho$ is the condensate density. The quasiparticle modes which
diagonalize $H_0$ interact via $H_{int}$. The terms
$A_{\vec{k,q}},B_{\vec{k,q}}$ are combinations of the $u$'s and
$v$'s originating from interference between different atomic
collision pathways, and are given by
\begin{eqnarray}\label{eq:A,B}
A_{\vec{k,q}}=2u_k\left(u_q u_{\abs{\vec{k-q}}}
 -u_q v_{\abs{\vec{k-q}}}
-v_q u_{\abs{\vec{k-q}}}\right)-2v_k\left(v_q v_{\abs{\vec{k-q}}}
 -v_q u_{\abs{\vec{k-q}}}
-u_q v_{\abs{\vec{k-q}}}\right)\\
B_{\vec{k,q}}=2u_k\left(v_q v_{\abs{\vec{k-q}}}
 -u_q v_{\abs{\vec{k-q}}}
-v_q u_{\abs{\vec{k-q}}}\right)-2v_k\left(u_q u_{\abs{\vec{k-q}}}
 -v_q u_{\abs{\vec{k-q}}}
-u_q v_{\abs{\vec{k-q}}}\right).
\end{eqnarray}
$A_{\vec{k,q}}$ and $B_{\vec{k,q}}$ are symmetric under the
exchange $\vec{q} \leftrightarrow \vec{k-q}$, as implied by Eq.
(\ref{eq:bel_ham}).  The different terms of $H_{int}$ describe
different processes resulting from the quasiparticle interactions.
These processes are described diagrammatically in Fig.
\ref{fig:arrows}. The first term in Eq. (\ref{eq:bel_ham})
 corresponds to the decay of one
quasiparticle into two, and is analogous to parametric down
conversion. This decay is referred to as Beliaev damping.  The
second term describes two quasiparticles colliding, and forming
one quasiparticle with the total momentum. This term governs the
so called Landau damping. The last two terms in $H_{int}$
(proportional to $B_{\vec{k',q'}}$) describe the spontaneous
formation and annihilation of three quasiparticles with zero total
momentum.

\subsubsection{Resonant wave mixing}
Consider the case where there are initially $N\ll N_0$ excitations
in the momentum mode $\vec{k}$, and no other excitations. The
state $\ket{N_{\vec{k}}}$ is degenerate under $H_0$ with the
states $\ket{(N-1)_{\vec{k}},1_{\vec{q'}},1_{\vec{k-q'}}}$ for
those $\vec{q'}$ that fulfill the resonance condition
\begin{equation}\label{eq:rescond}
\epsilon_{q'}+\epsilon_{\abs{\vec{k-q'}}}=\epsilon_k.
\end{equation}
The mode ${\vec{k}}$ can therefore decay via the Beliaev term in
$H_{int}$ into pairs of modes on a shell of a quasi-continuum of
modes conserving both energy and momentum. The decay rate $\Gamma$
is approximated by the Fermi golden rule (FGR) \cite{ours-3wm}:
\begin{equation}
\Gamma
=\frac{2\pi}{\hbar}\frac{g^{2}N_{0}N}{4V^{2}}\frac{1}{2}\sum
_{\mathbf{q}'} \abs{2A_{\mathbf{k},\mathbf{q}'}} ^{2}\delta (
\epsilon _{\mathbf{k}}-\epsilon _{\mathbf{q}'}-\epsilon
_{\mathbf{k}-\mathbf{q}'}). \label{eq:damping rate 1}
\end{equation}
The factors of 2 in Eq. (\ref{eq:damping rate 1}) are due to
bosonic exchange symmetry between $\vec{q'}$ and $\vec{k-q'}$. The
FGR is valid as long as the quasi-continuum contains enough modes
$N_{c}$ such that during the time of the experiment the average
population of these modes will be less than 1, i.e. $\Gamma t\ll
N_{c}$. The second (Landau) term in $H_{int}$ does not contribute
to the evolution at temperatures much lower than the chemical
potential, since there is only a negligible occupation of modes
other than $\vec{k}$. The third and forth terms  have no
contribution to the decay since they do not conserve energy.

\begin{figure}
  \centering
  \includegraphics[width=8cm]{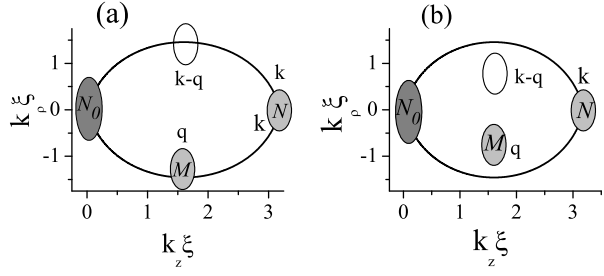}
  \caption{A schematic momentum-space description of three-wave mixing.
  (a) The $M$-fold occupation of the seed mode $\vec{q}$ fulfills the resonance condition of Eq. (\ref{eq:rescond}),
  and is therefore on the solid line which marks the quasi-continuum of possible states for Beliaev damping of mode
  $\vec{k}$.
  (b) The seed mode $\vec{q}$ is \emph{not} on the energy conserving shell, leading to non-resonant wave mixing.
  In both figures the condensate is marked as a dark gray ellipse,
   the initially occupied modes are marked as light gray ellipses,
   and the empty modes which fulfill momentum conservation are represented as empty ellipses.
   The shell of possible modes is calculated for $k\xi=3.2$,
    and is therefore not completely spherical as expected for atomic s-wave scattering, but rather lemon-shaped \cite{ours-beliaev}.
   Note that only in the high momentum limit do the plotted modes correspond to atomic momentum modes.  }\label{fig:schematic_shell}
\end{figure}

Now consider the evolution of the same state $\ket{N_{\vec{k}}}$,
when there is an initial seed of $M$ excitations of mode $\vec{q}$
which satisfies the resonance condition [Eq.(\ref{eq:rescond})],
i.e. it is on the energy-momentum conserving shell. Our initial
condition is thus $\ket{(N)_{\vec{k}},(M)_{\vec{q}}}$. Such a
situation is presented schematically in momentum space in Fig.
\ref{fig:schematic_shell}a. The condensate, which is the
quasiparticle vacuum is represented by a dark gray ellipse. The
initially occupied quasiparticle modes are plotted as light gray
ellipses, and the initially unpopulated mode $\vec{k-q}$ is
represented as an empty ellipse. The line describes the
quasi-continuum of possible modes for Beliaev damping from mode
$\vec{k}$. For $M\ll N_{c}$ the initially occupied mode $\vec{q}$
will merely enhance the decay rate in Eq. (\ref{eq:damping rate
1}) to
\begin{equation}
\Gamma_M=\Gamma+\frac{2\pi M}{2\hbar}\frac{g^{2}N_{0}N
A_{\vec{k,q}} }{V^{2}}.\label{eq:gammaM}
\end{equation}
 This enhancement of
$\Gamma$ is analogous to the optical power-broadening, where an
occupied mode of the electromagnetic field broadens the natural
linewidth of an atom $\Gamma$, which is also the decay rate due to
spontaneous emission \cite{API}. If, however $M> N_{c}$, a
situation that is experimentally achievable in cold atoms where
$N_{c}\sim10^3-10^4$, the dynamics are changed significantly from
the simple decay.  The term containing $\vec{k,q}$ in the sum will
dominate $H_{int}$ in Eq. (\ref{eq:bel_ham}) and  decay into two
unpopulated modes can be neglected, yielding $H_{int}\approx
H_{3W}^{res}$, where we define the resonant three-wave Hamiltonian
as
\begin{equation}\label{eq:Hint_no_det}
H_{3W}^{res}=\frac{g\sqrt{N_0}}{2V}\left[A_{\vec{k},\vec{q}}\left(\cre{b}{\vec{k}}\ann{b}{\vec{k-q}}\ann{b}{\vec{q}}+
  \ann{b}{\vec{k}}\cre{b}{\vec{k-q}}\cre{b}{\vec{q}}\right)\right].
\end{equation}
Since the modes $\vec{q}$ and $\vec{k-q}$ will both become
macroscopically occupied,  the second term in Eq.
(\ref{eq:Hint_no_det}) cannot be ignored. This case will be
studied in section \ref{sec:3wm}. Since the states
$\ket{(N-n)_{\vec{k}},(M+n)_{\vec{q}},n_{\vec{k-q}}}$ with
different $n$ are degenerate under $H_0$, the eigenstates of
$H_{3W}^{res}$ in this subspace are also eigenstates of $H$, and
once calculated, both dynamics and spectra are extracted in a
straightforward manner. In sections \ref{sec:3wm} and
\ref{sec:damping} $H_{3W}^{res}$ will lead to nonlinear
oscillations, and a splitting in the decay spectrum of the mode
$\vec{k}$.

\subsubsection{Non-resonant wave mixing}

The non-resonant terms of $H_{int}$ in Eq. (\ref{eq:bel_ham}) do
not contribute to the perturbative dynamics. They do, however,
lead to energy shifts. Using second order perturbation theory we
find an energy shift of the Bogoliubov quasiparticle energy of
mode $\vec{k}$ by
\begin{equation}
\delta\epsilon_k=\frac{g^2N_0}{4V^2}\wp\sum_{\vec{q'}}
\left[\frac{\abs{2A_{\vec{k,q'}}}^2}{2\left(\epsilon_k-\epsilon_q'-\epsilon_{\abs{\vec{k-q}}}\right)}
-\left(\frac{\abs{6B_{\vec{k,q'}}}^2}{2\left(\epsilon_k+\epsilon_q'+\epsilon_{\abs{\vec{k+q}}}\right)}\right)\right],\label{eq:lamb}
\end{equation}
where $\wp$ stands for the principal part. This energy shift is
analogous to the Lamb shift of atomic states due to the
Electromagnetic vacuum \cite{API}. In analogy to the resonant wave
mixing, we study the Hamiltonian (\ref{eq:bel_ham}), again with
the initial condition $\ket{(N)_{\vec{k}},(M)_{\vec{q}}}$. This
time we consider the case where the mode $\vec{q}$ is not on the
energy-momentum conserving shell defined by Eq.
(\ref{eq:rescond}), but is rather detuned by
$\hbar\Delta=\epsilon_{\vec{q}}+\epsilon_{\vec{\abs{k-q}}}-\epsilon_{\vec{k}}$,
as schematically shown in Fig. \ref{fig:schematic_shell}b. We
consider small values of detuning
$\hbar\Delta<<\epsilon_k,\epsilon_q$, hence we neglect the terms
$B_{\vec{k,q}}$ in Eq. (\ref{eq:bel_ham}) since the energy
denominators of the $B_{\vec{k,q}}$ terms are much larger than
those of the $A_{\vec{k,q}}$ terms. This is in analogy to the
rotating wave approximation in optics. To keep the states
$\ket{(N-n)_{\vec{k}},n_{\vec{k-q}},(M+n)_{\vec{q}}}$ degenerate
under $H_0$,  Eq. (\ref{eq:bog_ham}) is modified to the form
$H_0=E_g+\sum_{\vec{k'}}\epsilon_{k'}\cre{b}{\vec{k'}}\ann{b}{\vec{k'}}-\cre{b}{\vec{\abs{k-q}}}\ann{b}{\vec{\abs{k-q}}}\Delta$.
In order for $H$ to remain the same as in Eq. (\ref{eq:bel_ham})
we must generalize $H_{3W}^{res}$ to
\begin{equation}\label{eq:Hint_det}
H_{3W}=\frac{g\sqrt{N_0}}{2V}\left[A_{\vec{k},\vec{q}}\left(\cre{b}{\vec{k}}\ann{b}{\vec{q}}\ann{b}{\vec{k-q}}+
  \ann{b}{\vec{k}}\cre{b}{\vec{q}}\cre{b}{\vec{k-q}}\right)\right]+\cre{b}{\vec{\abs{k-q}}}\ann{b}{\vec{\abs{k-q}}}\Delta.
\end{equation}
We will show in section \ref{sec:damping} that $H_{int}$ leads to
a splitting in the damping spectrum of mode $\vec{k}$. The
detuning leads to an asymmetry in the peaks.  When the detuning is
large (but still smaller than $\epsilon_k/\hbar$) , one peak
vanishes, and the other peak dominates. The large peak is shifted
from $\epsilon_k$ due to the macroscopically occupied mode
$\vec{q}$ by
\begin{equation}
\delta \epsilon_k=
-\frac{g^{2}N_{0}M}{V^{2}}\frac{|A_{\mathbf{k},\mathbf{q}}|
^{2}}{2\Delta}. \label{eq:AC Stark shift}
\end{equation}
This shift in the spectrum is analogous to the ac-Stark shift and
appears in addition to the Lamb shift of Eq. (\ref{eq:lamb}).

\subsection{Wave mixing due to Bragg coupling}\label{sec:hamiltonians-2wm}
The interaction Hamiltonian described in Eq. (\ref{eq:bel_ham})
takes into account the  quantum depletion of atoms  from the
$\vec{k}=0$ mode in the many-body ground state of the BEC due to
interactions. In fact, this effect is what leads to the momentum
dependence of $A_{\vec{k,q}}$. However Eq. (\ref{eq:bel_ham}) is
derived under the assumption that at all times the population in
all modes much is smaller than $N_0$, i.e. the condensate is
undepleted. There are cases where the condensate is depleted, for
example by a strong two-photon Bragg coupling leading to Rabi
oscillations between momentum modes $0$ and $\vec{k_B}$
\cite{ours-dephasing}. The Bragg coupling is characterized by a
frequency $\omega_B$ and a wavevector $\vec{k}_B$, that are the
difference in the laser frequencies and wavevectors respectively
and by  an effective Rabi frequency $\Omega$ which describes the
coupling strength. The detuning of the Bragg pulse is
$\delta=\omega_B-\hbar k_B^2/2M$. In the rotating wave
approximation ($\delta\ll \hbar k_B^2/2M$) the resonant Bragg
pulse  couples the macroscopically occupied mode $\vec{k}=0$ to
$\vec{k}_B$, so we may estimate the effect of the Bragg pulse as
\cite{blakieBallagh}:
\begin{equation}\label{eq:Hbragg}
H_{2W}^{res}=
\frac{\hbar\Omega}{2}\left(\cre{a}{0}\ann{a}{\vec{k}_B}+\ann{a}{0}\cre{a}{\vec{k}_B}\right),
\end{equation}
where we have neglected coupling to all other momentum modes
($-\vec{k}_B,2\vec{k}_B,\ldots$)  which are far off resonance. So
far interactions are not included in this subsection, and
therefore we use the atomic basis rather than a quasiparticle
basis.
 As in Eq. (\ref{eq:Hint_det}), for a detuning from
resonance that is  much smaller than the detuning from other
momentum modes ($\abs{\delta}\ll\abs{\omega_B-\hbar (mk_B)^2/2M}$
for $m\neq 1$), the Bragg Hamiltonian is simply modified to
\begin{equation}\label{eq:Hbragg_det}
H_{2W}=
\frac{\hbar\Omega}{2}\left(\cre{a}{0}\ann{a}{\vec{k}_B}+\ann{a}{0}\cre{a}{\vec{k}_B}\right)-\hbar\delta\ann{n}{\vec{k}_B}.
\end{equation}
Here $\hat{n}_{\vec{k}}=\cre{a}{\vec{k}}\ann{a}{\vec{k}}$ is the
number operator of mode $\vec{k}$. Neglecting interactions between
the atoms, $H_{2W}$ describes the single particle two level
system. If one adds $H_{2W}$ to the Hamiltonian
(\ref{eq:H_mb_k_rep}), it is  no longer possible to use the
Bogoliubov approximation, since  the Bragg process depletes the
condensate. One can however include atomic interactions by taking
only the $\vec{k}=0$ and the $\vec{k}=\vec{k_B}$ modes into
account in the interaction term in Eq. (\ref{eq:H_mb_k_rep}). This
amounts to neglecting terms proportional to $v_k^2$ in the
Hamiltonian, such as the quantum depletion of the condensate.
Neglecting $v_k$ and approximating $u_k\approx 1$ is a good
approximation for momenta larger than the inverse of the healing
length $\xi=(8\pi\rho a_s)^{-1/2}$. In this regime the Bogoliubov
quasiparticle energy is a free particle parabola shifted by the
mean-field energy $\mu$, and the quasiparticle wavefunction is
equal to the free particle wavefunction \cite{Fetter-review}. The
coupling Hamiltonian including the resonant Bragg coupling and the
atomic
 interactions of  momentum modes $0$ and $\vec{k_{\vec{B}}}$ can be written as:
\begin{equation}\label{eq:Hbragg_int}
H_{2W}^{int}=
\frac{\hbar\Omega}{2}\left(\cre{a}{0}\ann{a}{\vec{k}_B}+\ann{a}{0}\cre{a}{\vec{k}_B}\right)
+\frac{g}{2V}\left[\ann{n}{0}\left(\ann{n}{0}-1)+\ann{n}{\vec{k}_B}\right(\ann{n}{\vec{k}_B}-1)+4\ann{n}{0}\ann{n}{\vec{k}_B}\right].
\end{equation}
Using the conservation of particles
$\ann{n}{0}+\ann{n}{\vec{k}_B}=N$, the interaction term in Eq.
(\ref{eq:Hbragg_int}) can be simplified to
$(N^2-N+2\ann{n}{0}\ann{n}{\vec{k}_B})g/2V$. For a small
excitation, i.e.  $ n_{\vec{k}_B}\ll n_0$, this coincides with the
mean-field energy $\ann{n}{\vec{k}_B}\mu$ of Bogoliubov
quasiparticles, while for a nearly depleted condensate, where
$n_0\ll n_{\vec{k}_B}$, the energy is the same due to the symmetry
$n_0\leftrightarrow n_{\vec{k}_B}$. This  situation describes a
condensate that is moving in the lab frame of reference with
momentum $\vec{k}_B$ and $n_0$ excitations with momentum
$-\vec{k}_B$ relative to the condensate. Here the mean field
energy per quasiparticle is $g n_{\vec{k}_B}/V$. $H_{2W}^{int}$ is
no longer a linear Hamiltonian and cannot be solved in first
quantization as a single particle problem. In fact, even the
Gross-Pitaevskii equation is inadequate for solving this
Hamiltonian for certain parameters \cite{ami}. The Hamiltonians
(\ref{eq:Hbragg}-\ref{eq:Hbragg_int}) are studied in section
\ref{sec:2wm}. Systems governed by Eq. (\ref{eq:Hbragg_int})
 also exhibit oscillations, a splitting in the excitation spectrum and a variation in the decay spectrum.
 These phenomena have been demonstrated experimentally
 \cite{ours-dephasing,ours-splitting}.

\section{Two-wave mixing}\label{sec:2wm}

In this section we discuss the diagonalization of the two-wave
mixing Hamiltonians (\ref{eq:Hbragg}-\ref{eq:Hbragg_int}). We
start with Eq. (\ref{eq:Hbragg}). Although this is a single atom
problem (since it includes no interactions) we diagonalize the
Hamiltonian in second quantization, in order to develop a notation
used later for problems such as three-wave mixing which are
many-body in nature. In subsection \ref{sec:rabi} we use this
simple notation to describe the perturbative dynamics of a
many-body problem including s-wave interactions between the atoms
oscillating between large momentum modes.

Although $H_{2W}$ involves two fields, the conserved quantity
$N=\cre{a}{0}\ann{a}{0}+\cre{a}{\vec{k_B}}\ann{a}{\vec{k_B}}$
allows us to diagonalize in a subspace with one quantum number
$n=n_{\vec{k_B}}$ ,varying between $0$ and $N$, and representing a
Fock state $\ket{n}=\ket{(N-n)_0,n_\vec{k_B}}$.

$H_{2W}$ couples between states in this subspace which have a
difference of $1$ in the quantum number $n$, and can be
represented in the resonant case by the $(N+1)\times{(N+1)}$
tri-diagonal matrix

\begin{equation}\label{eq:H2wMat}
H_{2W}^{res}=\frac{\hbar \Omega}{2}\left(\begin{array}{ccccc}
     0              & \sqrt{1(N-0)}     &  0              &  \cdots  & \\
    \sqrt{1(N-0)}   &            0      & \sqrt{2(N-1)}  &          & \\
     0              & \sqrt{2(N-1)}     & 0              & \ddots   & \\
      \vdots        &                   &\ddots            & \ddots & \sqrt{(N-0)1}\\
                    &                      &
                    &\sqrt{(N-0)1}& 0

\end{array}\right)\end{equation}

When $H_{2W}^{res}$ is diagonalized, we get a new set of $N+1$
eigenstates, that are dressed by the interaction, labelled by the
quantum numbers $m_x$, which can take $N+1$ different values.

\begin{figure}[tb]
\begin{center}
\includegraphics[width=8cm]{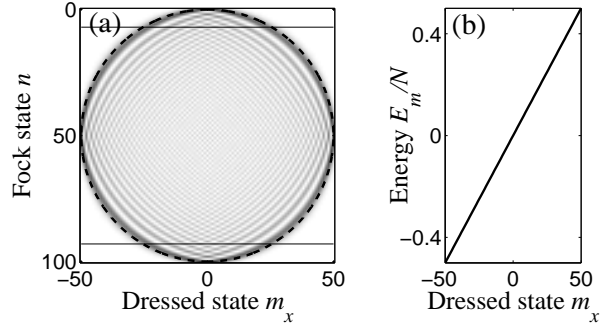}
\end{center}
\caption{(a) Absolute value squared of the transformation matrix
 between the Fock basis and the dressed basis which
diagonalizes $H_{2W}^{res}$ given in Eq. (\ref{eq:Hbragg})
$\abs{\braket{n}{m_x}}^2$, for $N=100$. Dashed line is the circle
$(n-j)^2+m_x^2=j^2$. The thin horizontal lines show the location
of the $n=7$ and $n=N-7$ rows which are plotted in Fig.
\ref{fig:two_wave_state}. (b) The spectrum of $H_{2W}^{res}$,
which is strictly linear. Energy is in units of $\hbar\Omega$
\label{fig:two_wave_mat}}
\end{figure}

 The squared absolute value of the matrix elements  of the transformation matrix
between the Fock states $\ket{n}$ and the dressed states $m_x$,
are shown in Fig. \ref{fig:two_wave_mat} for $N=100$, along with
the corresponding eigen-energies. As can be seen, the spectrum is
linear, $E_{m_x}=\hbar\Omega N m_x$, indicating this is indeed a
single particle problem. Once the transformation matrix is solved,
dynamics can be calculated trivially by projecting the initial
state on the dressed basis and evolving each dressed state
according to $\ket{m_x(t)}=e^{-iE_{m_x}t/\hbar}\ket{m_x(0)}$,
yielding the well known Rabi oscillations, with frequency
$\Omega$. These oscillations have been demonstrated experimentally
\cite{phillips-lattice}.

 \subsection{The Schwinger boson mapping}\label{sec:schwinger}
 We now turn to describing the problem in terms of angular
 momentum. For non-interacting resonant coupling the exact solution was found by Tavis and
 Cummings \cite{cummings68}.
 This approach will prove to be useful later when we
 deal with problems which are no longer linear.
 The two dimensional harmonic oscillator can be solved in different bases.
 Each oscillator can be solved separately and the eigenfunctions are
 tensor products of one dimensional oscillators.
 Another option is to
 work in polar coordinates, and find eigenstates of angular
 momentum and of the radial equation. The Schwinger boson mapping
 \cite{schwinger,sakurai}
 maps the Fock basis of decoupled one dimensional harmonic oscillators,
 with a fixed sum of excitations $N=n_0+n_\vec{k_B}$ onto the angular
 momentum basis with a fixed total angular momentum $j=N/2$. The
 mapping is defined by the following operators
\begin{eqnarray}
\ann{J}{}^+ & \equiv & \ann{a}{0}\cre{a}{\vec{k_B}}\label{eq:schwin_J+},\\
\ann{J}{}^- & \equiv & \ann{a}{\vec{k_B}}\cre{a}{0},\\
\ann{J}{z} & \equiv &
\frac{1}{2}\left(\cre{a}{\vec{k_B}}\ann{a}{\vec{k_B}}-\cre{a}{0}\ann{a}{0}\right).
\end{eqnarray}
Both representations fulfill the angular momentum commutation
relations, and this mapping is completed by associating the
quantum numbers
\begin{eqnarray}
m_z & \equiv & \frac{n_{\vec{k_B}}-n_{0}}{2},\\
j & \equiv &
\frac{n_{\vec{k_B}}+n_{0}}{2}=\frac{N}{2}\label{eq:schwin_j}.
\end{eqnarray}
 Thus a state
$\ket{(N-n)_0,n_\vec{k_B}}$ in the Fock basis is equivalent to the
state $\ket{j=N/2,m_z=n_\vec{k_B}-N/2}$ in the angular momentum
basis. Using this mapping we write the Hamiltonian in the angular
momentum basis simply as
\begin{equation}\label{eq:2wH_J}
H_{2W}^{res}=\frac{1}{2}\hbar\Omega(\ann{J}{}^+ +
\ann{J}{}^-)=\hbar\Omega \ann{J}{x}.\end{equation} The eigenstates
of $H_{2W}^{res}$, which we refer to as dressed states,  are
labelled by their projection on the $x$ axis $m_x=-j,\ldots,j$,
rather than their projection on the $z$ axis $m_z=n-j$, and the
spectrum is linear $E_{m_x}=\hbar\Omega m_x$. The transformation
matrix from the Fock basis to the dressed basis, shown in Fig.
\ref{fig:two_wave_mat}, is also straightforward to understand in
this representation, and is simply a N+1 dimensional
representation of an SO(3) rotation around the $y$ axis by an
angle $\theta=-\pi/2$ with matrix elements, known as the Wigner
matrix elements, $d^{j}_{m_x,m_z}(-\pi/2)$, which are defined
analytically in \cite{Wigner}.

In the limit of large $j$, we can use classical reasoning to
predict the shape of the transformation matrix. A classical spin
of length $j$ with a projection $m_z$ on the z axis could be
 found anywhere on a circle of radius $\sqrt{j^2-m_z^2}$  in the x-y
plane with equal probability. Therefore, the probability to
measure a certain projection $m_x$ on the x axis vanishes for
$\abs{m_x}>\sqrt{j^2-m_z^2}$ and is proportional to
$(j^2-m_z^2)^{-1/2}$ for $\abs{m_x}<\sqrt{j^2-m_z^2}$. For this
reason we expect that for large enough representations, the
transformation matrix should be bounded by the circle
$m_x^2+m_z^2=j^2$, and its amplitude should peak on this circle.
 In Fig. \ref{fig:two_wave_mat}(a), this circle is
plotted as a dashed line. Indeed, the largest amplitudes are found
on this circle, and the amplitudes outside are completely
negligible.

$H_{2W}^{res}$ is analogous to the Hamiltonian of a magnetic field
in the X direction applied to a spin j particle. The spectrum
presented in Fig. \ref{fig:two_wave_mat}(b) is therefore analogous
to the Zeeman energy of that particle.

\begin{figure}[tb]
\begin{center}
\includegraphics[width=8cm]{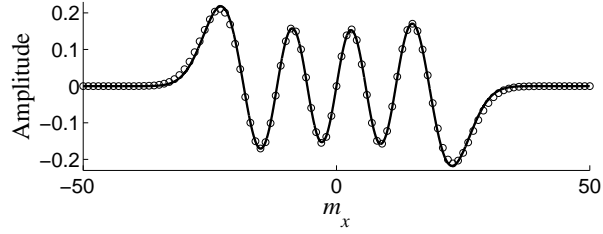}
\end{center}
\caption{The overlap between the Fock state $n=7$ and the dressed
state of $H_{2W}^{res}$ $m_x$,
$\braket{n=7}{m_x}=(-1)^{n-j-m_x}\braket{n=N-7}{-m_x}$ for
$N=100$. The harmonic oscillator approximation given by Eq.
(\ref{eq:matrix_sim_herm}) (line) is seen to be a good
approximation.} \label{fig:two_wave_state}
\end{figure}

If one of the momentum modes, say $0$, has a large occupation, the
operator
$\cre{a}{0}\ann{a}{\vec{k_B}}+\cre{a}{\vec{k_B}}\ann{a}{0}$ can be
estimated by $\sqrt{n_0}(\cre{a}{\vec{k_B}}+\ann{a}{\vec{k_B}})$
which is proportional to the position operator  for the mode
$\vec{k_B}$, whose eigenfunctions are the single harmonic
oscillator wavefunctions. Therefore for $n$ close enough to 0 the
representation of the Fock states in the dressed basis are indeed
very similar to the eigenfunctions of a single harmonic oscillator
and are given by \cite{rowe,beamsplitter}
\begin{equation}
\braket{n}{m_x}=d^j_{m_x,n-j}(\pi/2)\approx(-1)^{n} j ^{-1/4}
u_{n}\left( \sqrt{j} \arcsin \left(m_x/j\right)\right)
\label{eq:matrix_sim_herm} ,\end{equation} where $u_n(x)$ is the
$n$th eigenfunction of the harmonic oscillator. Figure
~\ref{fig:two_wave_state} presents the overlap of the Fock state
$n=7$ with the dressed basis $m_x$, $\braket{n=7}{m_x}$. The
overlap is seen to fit Eq. (\ref{eq:matrix_sim_herm}) remarkably
well. Using the symmetry $d^j_{m,m'}=(-1)^{m-m'}d^j_{-m,-m'}$ we
find that
$\left(-1\right)^{n-j-m_x}\braket{N-n}{-m_x}=\braket{n}{m_x}$ for
all $m_x$. This reflects the symmetry $0\leftrightarrow \vec{k_B}$
in $H_{2W}^{res}$.

\subsection{Detuning}\label{sec:detuning}
\begin{figure}[tb]
\begin{center}
\includegraphics[width=8cm]{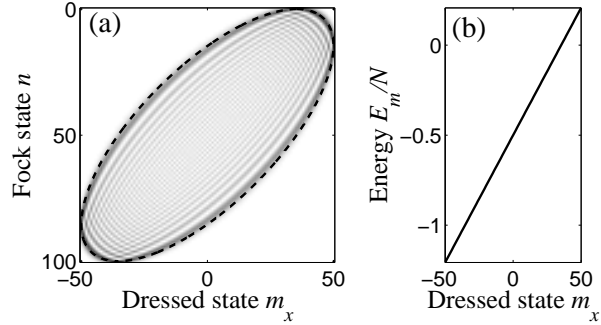}
\end{center}
\caption{(a) Absolute value squared of the transformation matrix
between the Fock basis and the dressed basis of $H_{2W}$,
$\abs{\braket{n}{m_x}}^2$, for $N=100$ and $\delta=\Omega$. Dashed
line is the shape $(n-j-(\delta/\Omega) m_x)^2+m_x^2=j^2$, which
is the generalization of the shape in Fig.
\ref{fig:two_wave_mat}(b) The spectrum of $H_{2W}$, which is again
strictly linear, with slope $\sqrt{\Omega^2+\delta^2}$, centered
at $-j\hbar\delta$. Energy is in units of $\hbar\Omega$.}
\label{fig:two_wave_det_mat}
\end{figure}

Next, we describe the case in which there is a detuning between
the Bragg frequency and the difference in the energy of the
momentum modes. We therefore add the detuning to $H_{2W}$ as we
did in Eq. (\ref{eq:Hbragg_det}). In the matrix (\ref{eq:H2wMat}),
a term $-n\hbar\delta$ is added to the $n$th element on the main
diagonal of the Hamiltonian. The matrix is diagonalized by the
transformation matrix shown in Fig. \ref{fig:two_wave_det_mat}
along with the spectrum of the Hamiltonian (\ref{eq:Hbragg_det})
for $N=100$ and $\delta=\Omega$. Again, the linear spectrum
reflects the single particle nature of the Hamiltonian.

In order to analytically solve Hamiltonian (\ref{eq:Hbragg_det})
we use the fact that
$\ann{n}{\vec{k_B}}=N/2+\ann{J}{z}=j+\ann{J}{z}$, to write
$H_{2W}$ in the angular momentum representation as
\begin{equation}\label{eq:two_wave_schwin_det}
H_{2W}=\hbar\Omega \ann{J}{x}-\hbar\delta
\left(\ann{J}{z}+j\right)=\hbar\sqrt{\Omega^2+\delta^2}\ann{J}{l}-j\delta.
\end{equation}
The energy spectrum is again linear, with a spacing of
$\hbar\sqrt{\Omega^2+\delta^2}$, centered about $-j\delta$. The
eigenstates are those of a spin component operator pointing in the
$l$ direction in the $x-z$ plane, the angle from the $z$ axis
being $\theta=\pi+\arctan\left(\Omega/\delta\right)$. Thus the
transformation matrix is once again a rotation by $\theta$ about
the $y$ axis. The shape of the transformation matrix is again
captured by classical reasoning (dashed line in Fig
\ref{fig:two_wave_det_mat}a).

Dynamics are also easily understood in  the angular momentum
picture, in terms of spin precession. The Fock basis is
characterized by a quantization axis pointing up on the Bloch
sphere, while the dressed basis has the quantization axis tilted
by an angle $\theta$ towards the $x$ axis. This gives a visual
understanding of the transformation matrix, as well as the
dynamics. The pseudospin evolves by precession about the axis $l$,
with frequency $\sqrt{\Omega^2+\delta^2}$. The well known
properties of Rabi oscillations are obtained  in a
straight-forward manner. Inversion can only be achieved when the
driving field is resonant to the transition between the two modes.
As the detuning increases, the amplitude of oscillation decreases
and the frequency, which is simply the energy difference between
dressed states increases. Since  we have neglected interactions,
the spectrum is strictly linear, and there is no dephasing.

As in the resonant case discussed in section \ref{sec:schwinger},
the symmetry of the rotation matrix allows us to relate the row
$n$ to $N-n$. For small $n$, both are approximated by properly
translated and stretched harmonic oscillator wavefunctions
\cite{rowe}.

\subsection{Rabi oscillations of interacting atoms}\label{sec:rabi}
Upon inclusion of interactions between the atoms, the problem
ceases to be a single body problem, and the second quantization
method reviewed in the previous sections becomes useful. The
inclusion of interactions in the resonant Bragg coupling of Eq.
(\ref{eq:Hbragg_int}) is correct in the limit $\vec{k_B}\xi>1$ as
discussed in section \ref{sec:hamiltonians-2wm}. In this
subsection we assume for simplicity $\delta=0$. Generalization to
non-resonant Bragg coupling is trivial.

\begin{figure}[tb]

\begin{center}
\includegraphics[width=8cm]{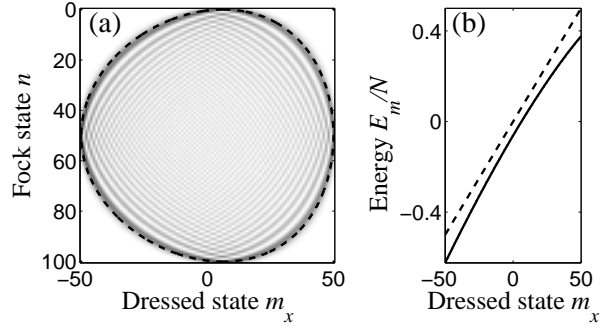}
\end{center}
\caption{(a) Absolute value squared of the transformation matrix
elements, for $H_{2W}^{int}$ with $\mu=\hbar\Omega/2$ and $N=100$.
The dashed line is the perturbative classical solution
$m_x=\pm\sqrt{j^2-m_z^2}+\mu m_z^2/4j$. (b) The spectrum (solid
line) is no longer linear. The non-interacting spectrum is shown
for reference as a dashed line. The first order perturbation of
Eq. (\ref{eq:2wmenergy}) is indistinguishable from the solid line.
Energy is in units of $\hbar\Omega$. } \label{fig:2wInt}
\end{figure}

In the angular momentum representation, the interaction term in
Eq. (\ref{eq:Hbragg_int}) can be rewritten up to a constant which
we neglect in the form
\begin{equation} \label{eq:H2w_int}
 H_{2W}^{int}=\hbar \Omega\ann{J}{x} + (\mu/N)\ann{J}{z}^2 .
 \end{equation}
The transformation matrix obtained by numerically diagonalizing
$H_{2W}^{int}$ is shown along with the obtained spectrum in
Fig.\ref{fig:2wInt} for $\mu=\hbar \Omega/2$.

The interactions are seen to change the circular shape of the
transformation matrix to an oval shape, breaking the symmetry
$m_x\leftrightarrow -m_x$. The spectrum which is no longer linear
indicates this is not a single particle problem.

In order to describe analytically the effect interactions have on
the dynamics, we use first order perturbation theory to calculate
the energy correction to the state $\ket{m_x}$. We start with the
eigen-energies of the non-interacting case
$E_{m_x}^{(0)}=\hbar\Omega m_x$, and add the correction
\begin{equation}
\label{eq:2WMInt}
E_{m_x}^{(1)}=\bra{m_x}H_{int}\ket{m_x}=-\bra{m_x}\frac{\mu}{N}\ann{J}{z}^2\ket{m_x}=
\frac{\mu}{2j}\bra{m_x}\frac{\ann{J}{x}^2-j(j+1)}{2}\ket{m_x}
\end{equation}
The overall energy of the state $\ket{m_x}$ is therefore
\begin{equation}
E_{m_x}=\mu\frac{j-1}{4}+\hbar\Omega
m_x+\frac{\mu}{4j}m_x^2.\label{eq:2wmenergy}\end{equation} This
perturbative result for the spectrum is indistinguishable from the
solid line in Fig. \ref{fig:2wInt}b which is obtained by exact
diagonalization. The dashed line in Fig. \ref{fig:2wInt}(a) is a
perturbative correction to that of Fig. \ref{fig:two_wave_mat}(a).
It is derived by assuming the energy is nearly linear, and
therefore $(\mu/\hbar\Omega)m_z^2/4j$ is added to $m_x$. This line
is seen to indeed describe the peak amplitudes of the
transformation matrix very well.

\begin{figure}[tb]
\begin{center}
\includegraphics[width=8cm]{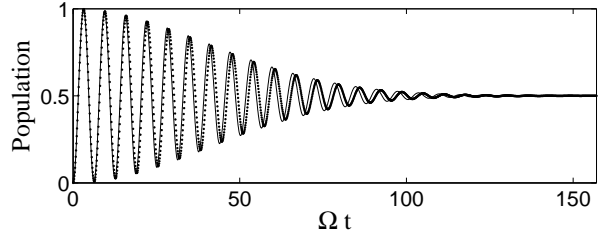}
\end{center}
\caption{The expectation value for population of mode $\vec{k_B}$
evolving according to $H{2W}^{int}$ with interactions for
$\mu=\hbar\Omega/2$, and $N=100$. The dots are according to the
numerical diagonalization of $\hbar
\Omega/2\ann{J}{x}+\mu/N\ann{J}{z}^2$. The line is according to
$H_{2W}$ with the first order correction to the energies
(\ref{eq:2wmenergy}). \label{fig:2wIntEvolution}}
\end{figure}
 The departure of the energy spectrum from a linear one, as shown
in Fig \ref{fig:2wInt}b, leads to a dephasing between the
different dressed states. According to Eq.
(\ref{eq:matrix_sim_herm}), an initial Fock state $n=0$ is a
gaussian with width $\sqrt{j}$. Hence a collapse should occur at a
time scale over which a $\pi$ phase is acquired in the quadratic
terms of the energies of the states $m_x=0$ and $m_x=\sqrt{j}$.
According to Eq. (\ref{eq:2wmenergy}) this energy is ${\mu}/{4}$,
leading to a collapse time of $t_{collapse}={4\pi\hbar}/\mu$. In
Fig \ref{fig:2wIntEvolution}, we see the decaying oscillations due
to the perturbative dynamics governed by Eq. (\ref{eq:2wmenergy})
fit the exact numerical solution remarkably well. A revival occurs
when each quadratic phase is a multiple of $2\pi$. This happens
for multiples of the $t_{revival}={4\pi \hbar N}/{\mu}$. In
realistic experimental conditions, $N$ is typically on the order
of $10^5$ and thus decoherence is likely to occur much before
$t_{revival}$ and we do not expect revivals to be seen.

The broadening in the spectrum which causes $t_{collapse}$ is
proportional to the chemical potential, and exists for a spatially
homogeneous system. It should not be confused with the broadening
in the resonance of Bogoliubov quasiparticles over a spatial
inhomogeneous BEC, known as inhomogeneous broadening \cite{line},
where the local density gives rise to a local chemical potential
and a local mean-field shift. For rapid Rabi oscillations this
well known inhomogeneous broadening is suppressed
\cite{ours-dephasing}. The slow decay discussed here is due to a
\emph{temporal} inhomogeneity, rather than a spatial inhomogeneity
in the mean-field energy.  The origin of this decay is in the
spread of dressed states spanning the initial state $m_z=-j$, each
having a slightly different Rabi frequency. The decay can be
explained in the Bloch sphere picture by a spread in the initial
conditions due to quantum uncertainty. The initial state $m_z=-j$
is classically described as a vector pointing down on a Bloch
sphere of radius $j$. However this is only the expectation value
of the angular momentum, and there is a fluctuation (independent
of $j$) in the initial value of $j_x$ and $j_y$. This uncertainty
in the initial conditions exists in the linear Rabi oscillation
problem too, but all vectors have the same oscillation frequency,
thus all possible vectors return to their initial value together
and there is no dephasing. Non-linearity causes the different
trajectories to have a slightly different Rabi frequency, thus
causing dephasing. This decay is therefore not predicted by the
homogeneous Gross-Pitaevskii equation, which corresponds to a
single trajectory of the expectation value of $\vec{j}$ on the
Bloch sphere.

\begin{figure}[tb]
\begin{center}
\includegraphics[width=8cm]{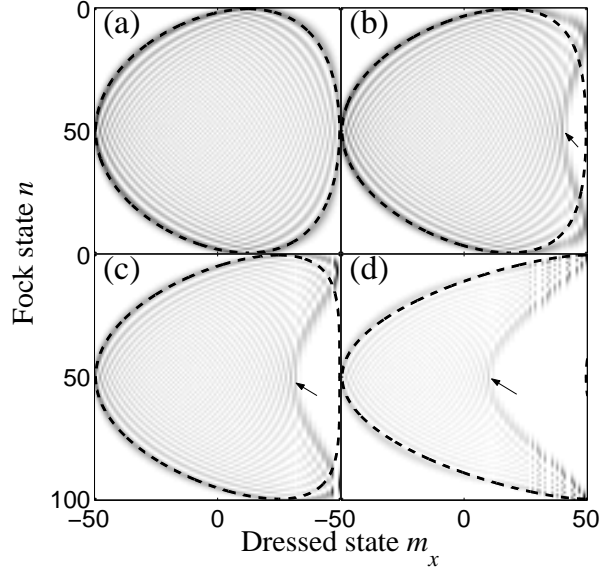}
\end{center}
\caption{The transformation matrices of $H{2W}^{int}$ for
different values of $\mu/\hbar\Omega=1$,$1.5$,$2$ and $4$ (a-d
respectively). For $\mu<\hbar\Omega$ the perturbative classical
equation $m_x=\pm\sqrt{j^2-m_z^2}+(\mu/\hbar\Omega) m_z^2/4j$
describes the region of high amplitude in the transformation
matrix quite well (dashed line). This approximation breaks down as
the instability comes in. At $\mu=2\hbar\Omega$, the turning point
reaches the initial dressed state vector, when starting from the
$n=0$ Fock state, and the dynamics depart from the perturbative
expectation (see Fig. \ref{fig:2WintDyn}) \label{fig:2WintMats}}
\end{figure}

The perturbative treatment discussed above  is correct for
$\mu<\hbar\Omega$.  It is known that for $\mu=\hbar\Omega$ a
dynamical instability occurs in the solution  of the two mode
Gross-Pitaevskii equation \cite{ami}. In our model the signature
of the instability is manifested by the deviation of the region of
high amplitude in the transformation matrix from the oval shape,
as seen in Fig. \ref{fig:2WintMats}.
For $\mu=\hbar\Omega$ the perturbative classical fit to the shape
of the transformation matrix is still good, and the curvature of
both vanish at $m_x=j$ (Fig. \ref{fig:2WintMats}a). For
$\mu>\hbar\Omega$ the transformation matrix becomes concave (Fig.
\ref{fig:2WintMats}b). The right hand side of the transformation
matrix varies greatly from the corresponding perturbative result,
while the left hand side is seen to follow the perturbative shape
even for $\mu>\hbar\Omega$.

We find numerically that for $m_x$ larger than a critical
$m_x^{c}$, the energy is doubly degenerate, with one corresponding
eigenvector located only in the top half of the matrix ($n<j$ or
$m_z<0$), and the other located in the bottom of the matrix ($n>j$
or $m_z>0$). The value of $m_x^{c}$ is found to decrease towards
the turning point in the transformation matrix (arrows in Fig.
\ref{fig:2WintMats}b-d) as the matrix size $N$ is increased. Thus
we conclude that in the thermodynamic limit $m_x^{c}$ coincides
with the turning point in the shape of the matrix.

To the left of the instability, eigenvectors have nonvanishing
amplitudes both in the upper and the lower half of the
transformation matrix. An initial state that spans a number of
such eigenvectors is able to perform Rabi oscillations. The
accumulated phases cause alternating constructive and destructive
interferences for a large range of $n$. To the right of the
instability this is not the case. In the thermodynamic limit, each
eigenvector is located completely either in the top half or in the
bottom half of the matrix. Therefore all eigenvectors spanned by a
state in the upper half will always interfere in the upper half
and $n>j$ will never be reached. Thus in this region inversion is
impossible, an effect known as macroscopic self-trapping
\cite{smerzi}.

\begin{figure}[tb]
\begin{center}
\includegraphics[width=8cm]{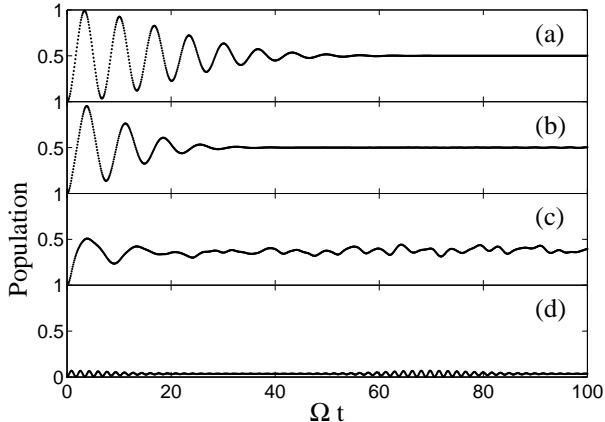}
\end{center}
\caption{The expectation value for  population of state
$\vec{k_B}$ as a function of time for the same parameters as in
fig \ref{fig:2WintMats}. For $\mu<2\hbar\Omega$ the dynamics are
described quite well by the perturbative treatment (a,b). As $\mu$
approaches $2\hbar\Omega$, the turning point in Fig.
\ref{fig:2WintMats}(c) reaches the initial dressed states and the
dynamics become unstable (c). For $\mu>2\hbar\Omega$ (d), we find
weak oscillations which cannot reach inversion (macroscopic
self-trapping). \label{fig:2WintDyn}}
\end{figure}

The Fock state $n=0$ is approximated by a Gaussian superposition
of dressed states of width $\sqrt{j}$, centered at
$(\mu/\hbar\Omega) j/4$. For $\mu=2\hbar\Omega$, the point of
instability is found to reach $\mu j/4$. Thus as $\mu$ approaches
$2\Omega$, the point of dynamic instability reaches the trajectory
of the initial Fock state $n=0$. Fig. \ref{fig:2WintDyn} shows the
time dynamics of the initial condition $n=0$ for the same ratios
of $\mu/\hbar\Omega$ as in Fig. \ref{fig:2WintMats}. For
$\mu<2\hbar\Omega$, we see Rabi oscillations. Even though there is
a turning point (in Fig. \ref{fig:2WintMats}b), as long as it is
not on the eigenstates occupied by $n=0$, oscillations are
regular. For $\mu\approx 2\hbar\Omega$ (Fig.
\ref{fig:2WintMats}c), the oscillations decay rapidly, and there
is a noisy fluctuation with no clear frequency in the population
of $n$. Finally, for $\mu>2\hbar\Omega$ we see the macroscopic
self-trapping (Fig. \ref{fig:2WintMats}d).

In the next section we move on to describe three-wave mixing with
the tools developed in this section. The three-wave Hamiltonian
derived in section \ref{sec:hamiltonians-3wm} describes
interactions also in the low $k$ limit, but are limited to weak
excitations which do not deplete the macroscopically occupied
mode.

\section{Three-wave mixing}\label{sec:3wm}

In this section we discuss three-wave mixing of Bogoliubov
quasiparticles, as governed by the Hamiltonian in Eq.
(\ref{eq:Hint_det}). In the limit of large momentum, where the
quasiparticle operators $\ann{b}{\vec{k}}$ coincide with the
atomic operators $\ann{a}{\vec{k}}$, the three-wave mixing
described here corresponds to  four-wave mixing of atomic matter
waves \cite{band-4wm}. The fourth wave  is the condensate which is
the vacuum of the Bogoliubov quasiparticles.  In the limit of low
momentum, the four-wave picture is incorrect, as each
quasiparticle mode $\ann{b}{\vec{k}}$ is a superposition of atomic
operators with momenta $\hbar\vec{k}$ and $-\hbar\vec{k}$. This
Hamiltonian does not conserve the total number of excitations
$n_\vec{k}+n_\vec{q}+n_\vec{k-q}$, indicating these excitations
are not particles, but rather quasiparticles whose number need not
be conserved.  There are two independent conserved quantities
$M=n_\vec{q}-n_\vec{k-q}$, and $N=n_\vec{k}+n_\vec{k-q}$, which
lead to the possibility of diagonalizing a subspace of Eq.
(\ref{eq:Hint_det}) with only one quantum number $n=n_\vec{k-q}$
representing the Fock state
$\ket{(N-n)_\vec{k},(M+n)_\vec{q},n_\vec{k-q}}$, where $n$ varies
between $0$ and $N$. The fact that the difference in the
occupation of modes $\vec{q}$ and $\vec{k-q}$ is constant can be
used as a source of number-squeezed states. In the low momentum
regime suggested here, dephasing is slow and pair correlated
atomic beams should be easier to generate than in the high
momentum regime studied experimentally in \cite{ketterle-4wm}.

\subsection{Resonant three-wave mixing}\label{sec:res3wm}
We consider first the resonant case $\Delta=0$ of Eq.
(\ref{eq:Hint_no_det}), where the sum of energies of the
quasiparticle modes $\vec{q}$ and $\vec{k-q}$ is equal to the
energy of mode $\vec{k}$, i.e.
$\epsilon_k=\epsilon_\vec{q}+\epsilon_{\abs{\vec{k-q}}}$. The
modes $\vec{q}$ and $\vec{k-q}$ are thus on the energy-momentum
conserving surface for Beliaev damping from mode $\vec{k}$ as
shown in Fig. \ref{fig:schematic_shell}a. $H_{3W}$ couples between
pairs of modes in this subspace which have a difference of $1$ in
$n$, and can be represented by the $(N+1)\times{(N+1)}$
tri-diagonal matrix

\begin{equation}\label{eq:H3wMat}
H_{3W}^{res}=\frac{K}{2}\left(\begin{array}{ccccc}
     0              & \sqrt{(N-0)M}     &  0              &  \cdots  & \\
    \sqrt{(N-0)M}   &            0      & \sqrt{(N-1)2(M+1)}  &          & \\
     0              & \sqrt{(N-1)2(M+1)}     & 0              & \ddots   & \\
      \vdots        &                   &\ddots            & \ddots & \sqrt{N(M+N)}\\
                    &                      &
                    &\sqrt{N(M+N)}& 0

\end{array}\right)\end{equation}
with a coupling term $K=g\sqrt{N_0}A_{\vec{k,q}}/V=\mu
A_{\vec{k,q}}/\sqrt{N_0}$. When $H_{3W}^{res}$ is diagonalized, we
get a new set of $N+1$ eigenstates, dressed by interactions,
$\ket{m_x} $, where $m_x$ varies between $-N/2$ and $N/2$.

\begin{figure}[tb]
\begin{center}
\includegraphics[width=8cm]{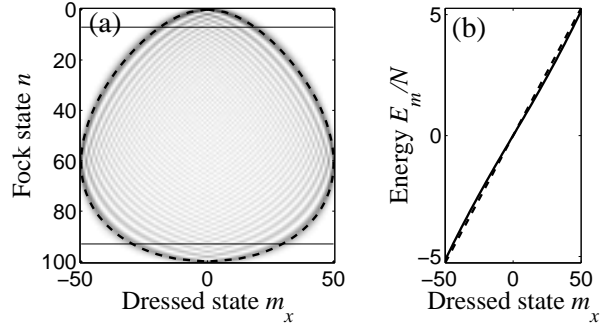}
\end{center}
\caption{(a) Absolute value squared of transformation matrix of
$H{3W}^{res}$ (Eq. (\ref{eq:Hint_no_det})) for $N=100, M=50$ and
$\Delta=0$. The thin lines are at the rows $n=7$ and $N-7$. The
dashed line is the line
$m_x^2+m_z^2\times\left(M+n\right)\left(M+\bar{n}\right)=j^2$.(b)
The exact spectrum (solid line) of $H_{3W}^{res}$ linearly
approximated by $\sqrt{M+\bar{n}}\, m_x$ (dashed line). Energy is
in units of $K$. The second order perturbation approximation (Eq.
\ref{eq:epert3w}) is indistinguishable from the solid line.
\label{fig:3wm}}
\end{figure}

 The
square of the transformation matrix elements are shown in Fig.
\ref{fig:3wm} for $N=100, M=50$
 along with the eigen-energies. As
can be seen by comparison with Fig. \ref{fig:two_wave_mat}, the
spectrum and the transformation matrix seem similar to those of
$H_{2W}^{res}$. In fact, as the seed $M$ is taken to be larger
than $N$, the factor $\sqrt{M+n}$ becomes more uniform, and can be
taken out of the matrix (\ref{eq:H3wMat}) as a multiplicative
factor. Thus in the limit $M\gg N$, both transformation matrix and
eigenvalues are well approximated by the two-wave mixing solution
described in section \ref{sec:2wm}, with $\hbar\Omega\approx
K\sqrt{M}$.

Even when $M$ is smaller than $N$ (but still large enough to seed
mixing), the factors originating from the $\vec{k-q}$ amplitudes
in adjacent matrix elements of $H_{3W}$, are almost the same,
$\sqrt{M+n}\approx\sqrt{M+n+1}$. We use this  to approximate Eq.
(\ref{eq:H3wMat}) \emph{locally} as a two-wave mixing Hamiltonian
multiplied by a local $\sqrt{M+n}$.

The transformation matrix in Fig. \ref{fig:3wm}(a) is indeed very
well approximated by stretching the  $n$'th line of the rotation
matrix (Fig. \ref{fig:two_wave_mat}(a)) by a factor
$\sqrt{M+n}/\sqrt{M+\bar{n}}$. $\bar{n}$ is the $n$ for which the
derivative of $\sqrt{N-n}\sqrt{M+n}\sqrt{n}$ with respect to $n$
vanishes. For $M\gg N$ the approximation of
$H_{3W}^{res}=H_{2W}^{res}$ becomes exact.

\begin{figure}[tb]
\begin{center}
\includegraphics[width=8cm]{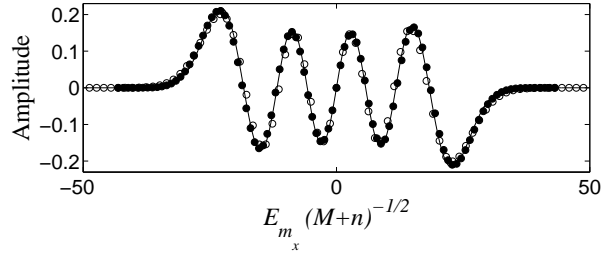}
\end{center}
\caption{The $n=7$ (filled circles) and the $n=N-7$ (empty
circles) rows of the transformation matrix presented in Fig.
\ref{fig:3wm}(a), divided by $\sqrt{M+n}$. The x axis is the
energy as calculated by Eq. (\ref{eq:epert3w}), and divided by
$\sqrt{M+n}$. The amplitudes of the different Fock states are
normalized by $[(M+n)/(M+\bar{n})]^{1/4}$. The line is the same
function (Eq. \ref{eq:matrix_sim_herm}) as in the two-wave mixing
in Fig. \ref{fig:two_wave_state}. \label{fig:3ws}}
\end{figure}

As can be seen in Fig. \ref{fig:3ws}, for small $n$, both rows $n$
and $N-n$ agree remarkably well with the function
(\ref{eq:matrix_sim_herm}) when plotted versus the spectrum of
$H_{3W}^{res}$, as in the case of two-wave mixing, even for $M<N$.

The spectrum is nearly linear centered around 0 with a slope
$dE=K\sqrt{M+\bar{n}}$. This leads to oscillations, with
oscillation period of $t=h/dE$ in analogy to the two-wave mixing
case. These oscillations decay due to the fact that the spectrum
is not strictly linear. In order to model analytically the
nonlinearity in the spectrum, we replace both operators
$\ann{b}{\vec{q}}$ and $\cre{b}{\vec{q}}$ by the operator
$\sqrt{M+\cre{b}{\vec{k-q}}\ann{b}{\vec{k-q}}}$ where the square
root of an operator is determined by a Taylor expansion. Since in
the Hamiltonian (\ref{eq:Hint_no_det}) $\ann{b}{\vec{q}}$ commutes
with $\cre{b}{\vec{k}}\ann{b}{\vec{k-q}}$, we approximate Eq.
(\ref{eq:Hint_no_det}) in a symmetric form:
\begin{equation}\label{eq:Hpert_3W_bs}
H_{3W}^{res}\approx \frac{K}{2}\left(
\sqrt{M+\cre{b}{\vec{k-q}}\ann{b}{\vec{k-q}}}\,
\cre{b}{\vec{k}}\ann{b}{\vec{k-q}}+\ann{b}{\vec{k}}\cre{b}{\vec{k-q}}\sqrt{M+\cre{b}{\vec{k-q}}\ann{b}{\vec{k-q}}}\right).
\end{equation}
We now define angular momentum operators in complete analogy to
Eqs. (\ref{eq:schwin_J+}-\ref{eq:schwin_j}) with the substitution
$\ann{a}{\vec{k_B}}\rightarrow\ann{b}{\vec{k-q}}$ and
$\ann{a}{0}\rightarrow\ann{b}{\vec{k}}$. In the angular momentum
representation Eq. (\ref{eq:Hpert_3W_bs}) becomes
\begin{equation}\label{eq:Hpert_3W}
H_{3W}^{res}\approx \frac{K}{2}\left(
\sqrt{M+j+\ann{J}{z}}\,\ann{J}{}^++\ann{J}{}^-\sqrt{M+j+\ann{J}{z}}\right).
\end{equation}
To the lowest order in $\ann{J}{z}/(M+j)$ we have
$H_{3W}^{res}\approx\hbar \bar{\Omega}\ann{J}{x}$ which is simply
$H_{2W}^{res}$ in Eq. (\ref{eq:2wH_J}), where
$\hbar\bar{\Omega}=K\sqrt{M+j}$. We use perturbation theory to
second order in $\ann{J}{z}/(M+j)$, to find the corrections to the
energy due to the non-linearity.

\begin{equation}
E_{m_x}=\hbar\bar{\Omega}\left[\left(1-\frac{j(j+1)}{16(M+j)^2}\right)m_x+5\frac{m_x^3}{16(M+j)^2}\right],\label{eq:epert3w}
\end{equation}
The perturbative spectrum obtained in Eq. (\ref{eq:epert3w}) is
indistinguishable from the solid line in Fig. \ref{fig:3wm}(b).

\begin{figure}[tb]
\begin{center}
\includegraphics[width=8cm]{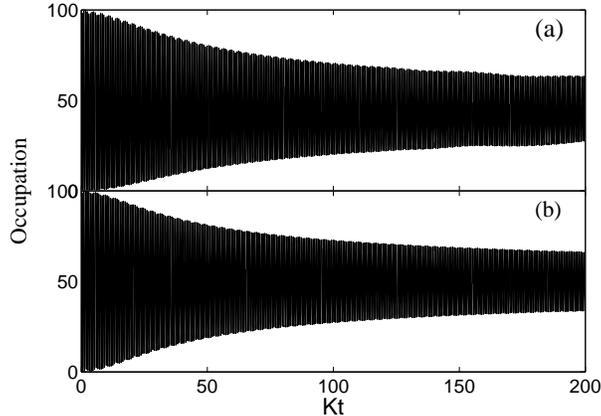}
\end{center}
\caption{The expectation value of the occupation of mode
$\vec{k-q}$ as a function of time obtained by  numerical
diagonalization (a) and perturbative calculation (b) of
$H_{3W}^{res}$, for the same parameters as in Fig. \ref{fig:3wm},
and starting with the Fock state $n=0$. Time is in units of
$\hbar/K$. } \label{fig:3wm_dynamics}
\end{figure}

As in the nonlinear case studied in section \ref{sec:rabi}, the
nonlinearity in the spectrum leads to dephasing in the dynamics.
Fig. \ref{fig:3wm_dynamics} compares between the numerical (a) and
the perturbative (b) results of the population of mode $\vec{k-q}$
vs. time, with the initial condition $n_{k-q}=0$. It is seen that
the perturbative solution captures the non-linear dynamics quite
well even for $M<N$. The rapid oscillation at frequency $dE/h$ are
unresolved in Fig. \ref{fig:3wm_dynamics}, but the envelope is
seen to be in good agreement. As expected from wave mixing, there
are oscillations between the population of the quasiparticle modes
$\vec{k}$ and $\vec{k-q}$. The seed in mode $\vec{q}$ oscillates
as well and has the occupation $M+n$. The initial Fock state is a
gaussian superposition of dressed states of width
$\sqrt{jM/(M+\bar{n})}$, thus the decay time is approximated
analytically by $t_{decay}=h
/E_{\sqrt{jM/(M+\bar{n})}}\sim\frac{32\pi\sqrt{j}}{5\bar{\Omega}}$
in agreement with Fig. \ref{fig:3wm_dynamics} . There is a revival
after time $t_{revival}\propto N$, due to the finite number of
quasiparticles. This revival time however is much longer than the
decoherence time for realistic values of $N$. Note that as opposed
to the interacting Rabi oscillating atoms of section
\ref{sec:rabi}, in which oscillation period and decay time are
independent of $N$, both times obtained from Eq.
(\ref{eq:epert3w}) scale as $\sqrt{j}$, when  the ratio $M/2j$ is
kept constant. Solving for a large system of size $j$, would
require the diagonalization of matrices of size $2j+1$. An
approximate solution therefore can be attained by solving for
moderate $j'$ and scaling the result by $\sqrt{j/j'}$.

\subsection{Detuning}\label{sec:3wm_det}
In the previous subsection, the seed, mode $\vec{q}$, was on the
energy-momentum conserving surface of mode $\vec{k}$. Exact energy
conservation is not a necessary condition for wave mixing as also
a populated mode with a small detuning $\Delta$ from this surface,
as shown in Fig. \ref{fig:schematic_shell}b will lead to wave
mixing. Figure \ref{fig:spectrum vs. detuning} shows the dressed
state spectrum obtained by numerical diagonalization of
$H_{3W}^{res}$ (Eq. (\ref{eq:Hint_det})) at different detunings
for $N=10$ and $M=5$.

When a detuning term is added to $H_{3W}$, Eq. (\ref{eq:Hpert_3W})
is modified to
\begin{equation}\label{eq:Hpert_3W_det}
H_{3W}\approx\frac{K}{2}\left(\sqrt{M+j+\ann{J}{z}}\ann{J}{}^++\ann{J}{}^-\sqrt{M+j+\ann{J}{z}}\right)
+\hbar\Delta\left(j+\ann{J}{z}\right).
\end{equation}
A perturbative approximation to the energy spectrum is obtained in
a similar fashion to Eq. (\ref{eq:epert3w}), with the Hamiltonian
(\ref{eq:Hpert_3W_det}) and is indistinguishable from the solid
lines in Fig. \ref{fig:spectrum vs. detuning}. At detunings much
larger than $K/\hbar$, the Hamiltonian (\ref{eq:Hpert_3W_det}) is
nearly diagonalized in the Fock basis $n$ or $m_z=n-j$. The energy
spacing, and therefore the energy span of the spectrum, increases
linearly with the detuning. At large negative detunings, the
lowest energy dressed state corresponds to the Fock state where
all of the excitations are in modes $\vec{k}$ and $\vec{q}$,
whereas at large positive detunings the lowest energy dressed
state corresponds to the Fock state where all of the excitations
are in modes $\vec{q}$ and $\vec{k-q}$. In analogy to rapid
adiabatic passage between two atomic levels, using a sweep in the
laser frequency, an adiabatic transfer of the excitations
population between the excitation modes of a BEC is possible using
a sweep in the detuning. Given an initial state where all of the
excitations in modes $\vec{k}$ and $\vec{q}$ at a large negative
detuning, an adiabatic sweep of the detuning to a large positive
value, would adiabatically transfer all of the excitations into
modes $\vec{q}$ and $\vec{k-q}$. A possible way to sweep the
detuning during the experiment would be the use of a magnetic
Feshbach resonance to tune $g$ \cite{Cornell_Feshbach}.

\begin{figure}[tb]
\begin{center}
\includegraphics[width=8cm]{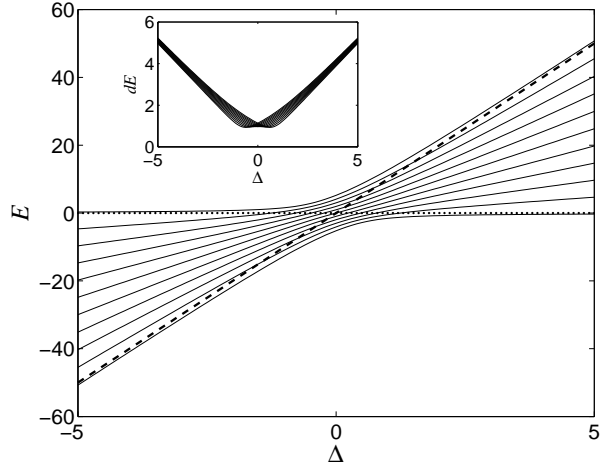}
\end{center}
\caption{ The spectrum of $H_{3W}$ for $N=10$ and $M=5$ vs.
detuning.
 Detuning is in units of $\bar{\Omega}$, and
energy in units of $\hbar\bar{\Omega}=K\sqrt{M+j}$. The
perturbative result is indistinguishable from the exact
calculation (solid line). Dashed line is
 $E=N\hbar\Delta$, the lowest (largest) energy for large negative
 (positive) detuning. Dotted line is $E=0$.
 For large negative detuning, the lowest dressed state
corresponds to the initial state $n=N$, while for large positive
detuning, the lowest dressed state corresponds to $n=0$. By
adiabatically sweeping the detuning, one can transfer all
population from modes $\vec{k}$ and $\vec{k-q}$ to modes $\vec{q}$
and $\vec{k-q}$. The inset is the difference between adjacent
eigen-energies. Non-linearity initially increases  with
$\abs{\Delta}$, reaches a maximum which depends weakly on $M/N$
and is always at $\abs{\Delta}\approx \bar{\Omega}$, and then
decreases for large $\abs{\Delta}$. } \label{fig:spectrum vs.
detuning}
\end{figure}

In analogy with  Eq. (\ref{eq:Hbragg_det}), we find that to zero
order in $\ann{J}{z}/(M+j)$, $H_{3W}$ in Eq.
(\ref{eq:Hpert_3W_det}) is given by
$H_{3W}^{(0)}=\hbar\sqrt{\bar{\Omega}^2+\Delta^2}\ann{J}{l}$,
where the axis $l$ points in an angle
$\theta=\pi+\arctan(\bar{\Omega}/\Delta)$ from the $z$ axis in the
$x-z$ plane.  For nonzero detuning the first order perturbative
correction to the dressed states' energies is proportional to
$\ann{J}{z}/(M+j)$. Therefore as the detuning is increased from
zero, the spread in the differences between adjacent dressed
states energies (non-linearity in the spectrum) is expected to
grow. On the contrary, for large detuning Eq.
(\ref{eq:Hpert_3W_det}) is dominated by the last term and
$H_{3W}^{(0)}$ is proportional to $\abs{\Delta}$, so the spectrum
approaches the linear spectrum of two-wave mixing with large
detuning. This is simply the zeeman shift in the angular momentum
representation. We therefore expect there will be an intermediate
value of detuning for which the spread in energy differences will
be maximal. The inset of Fig. \ref{fig:spectrum vs. detuning}
shows the energy difference between adjacent dressed states for
the same parameters as in the main figure. The non-linearity of
the spectrum indeed increases initially with the detuning, due to
the quadratic nonlinearity which is added in the presence of
detuning, and decreases towards $\Delta$ for large $\abs{\Delta}$.

\begin{figure}[tb]
\begin{center}
\includegraphics[width=8cm]{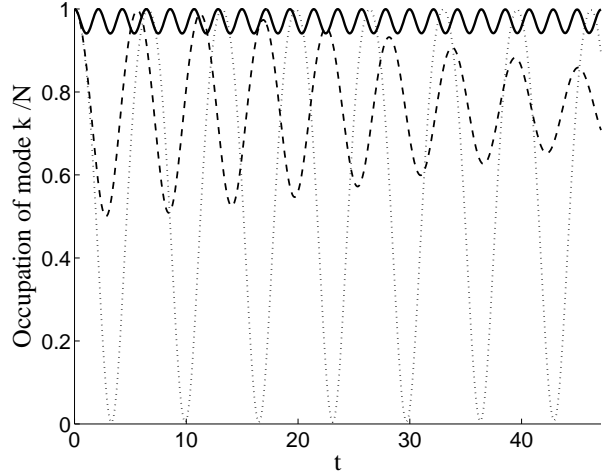}
\end{center}
\caption{The expectation value of the occupation  of mode
$\vec{k}$ as a function of time, for three different detuning
values, $\Delta=0$, $\Delta=\bar{\Omega}$ and
$\Delta=3\bar{\Omega}$ (dotted, dashed and solid lines
respectively) for $N=100$ and $M=50$. The oscillation decay time
for $\Delta=\bar{\Omega}$ is much shorter than for $\Delta=0$ and
$\Delta=3\bar{\Omega}$, in agreement with the  energy differences
in the dressed state spectra for the these detunings (see inset of
Fig. \ref{fig:spectrum vs. detuning}). Time is in units of
$1/\bar{\Omega}$. } \label{fig:oscillations}
\end{figure}

 The spread in energy differences is manifested  as dephasing in
the dynamics of the system. Figure \ref{fig:oscillations} shows
$\left<\cre{b}{\vec{k-q}}\ann{b}{\vec{k-q}}\right>$ as a function
of time, for three different detuning values, $\Delta=0$ (dotted
line), $\Delta=\bar{\Omega}$ (dashed line) and
$\Delta=3\bar{\Omega}$ (solid line), beginning from the initial
state $\ket{n_\vec{k}=N,n_\vec{q}=M,n_\vec{k-q}=0}$ with $N=100$
and $M=50$. The oscillation frequency is found to increase with
detuning to a value of
$\Omega_{eff}\equiv\sqrt{\bar{\Omega}^{2}+\Delta^{2}}$. The
oscillations amplitude is seen to decrease with larger detuning as
$1/\Delta^{2}$. The decay time of oscillations initially
increases, and then decreases, in agrement with the spread in the
energy differences plotted in the inset of Fig. \ref{fig:spectrum
vs. detuning}, for which the maximal spread is for
$\Delta/\bar{\Omega}\approx 1.5$. We use perturbation theory, in
analogy to Eq. (\ref{eq:Hpert_3W}), to calculate the spectrum and
$t_{decay}$ in the presence of detuning. These perturbative
results are found to fit the exact solution very well.

\section{Beliaev Damping}\label{sec:damping}

Thus far we have discussed the time evolution of the system within
the dressed state subspace, which is defined by the initial
population of the $\vec{k}$ and $\vec{q}$ modes, and have
neglected damping into empty modes.  In analogy to the treatment
of spontaneous photon scattering as a transfer between dressed
state subspaces of an atom interacting with a laser field
\cite{API}, we now consider scattering into empty modes as
transfer between dressed state subspaces. The damping of
excitations from the mode $\vec{k}$ is no longer elastic, but
rather carries the energy difference between the dressed states
among which it occurred. The extra energy is taken from the
interaction between the wave packets in the $\vec{k}$ and
$\vec{q}$ modes and was initially put into the system by the
lasers which excited the condensate into the $(N,M)$ subspace. We
consider only damping from the momentum mode $\vec{k}$ since the
damping rate from momentum modes $\vec{q}$ and $\vec{k-q}$ is
typically much smaller \cite{ours-beliaev}.

The damping  is an irreversible process due to the quasi-continuum
of  $N_c$ modes $\vec{q}'$ (other than $\vec{q}$ and $\vec{k-q}$)
on the energy-momentum conserving surface. The transition
amplitudes between all possible pairs of dressed states  under
\begin{equation}
H_{damping}=\frac{g}{2V}\sqrt{N_{0}}\sum_{\vec{q'}\neq\vec{q,k-q}}
A_{\vec{k,q'}}\left(\cre{b}{\vec{q'}}\cre{b}{\vec{k-q'}}\ann{b}{\vec{k}}\right),\label{eq:Hdamping}
\end{equation}
 reveal the spectral
structure of the damping process.

\begin{figure}[tb]
\begin{center}
\includegraphics[width=8cm]{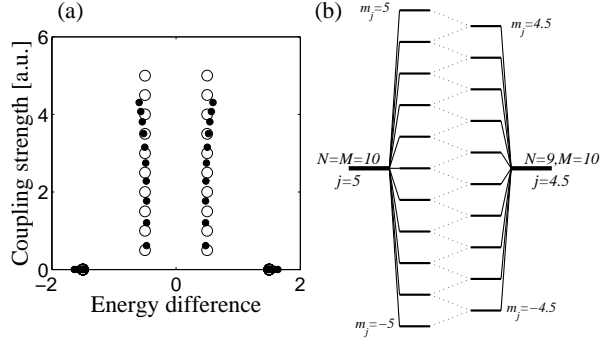}
\end{center}
\caption{ (a) The strength of coupling between pairs of dressed
states in the two subspaces $N=10,M=10$ and $N'=9,M=10$ as a
function of the energy difference between the two states. In the
linear two-wave approximation (empty circles) the coupling is
strictly only to the two states closest in energy, and the
coupling strength is linear in $m_x$ as implied in  Eq.
(\ref{eq:Hdecay}). For three-wave mixing (filled circles), there
is a negligible coupling to other states, and the coupling is no
longer linear in $m_x$.(b) Schematic drawing of Beliaev damping as
transfer between the subspaces.} \label{fig:Schematic transfer}
\end{figure}

The decay from mode $\vec{k}$ decreases $N$ by 1 and does not
change the value of $M$, Hence in the angular momentum
representation, a dressed state $\ket{j,m_x}$ can decay into
$\ket{j-1/2,m'_x}$. The matrix elements of the Hamiltonian
$H_{damping}$ between numerically obtained eigenstates of $H_{3W}$
(Eq. \ref{eq:Hint_det}) $\bra{j-1/2,m_x'}H_{damping}\ket{j,m_x}$,
are plotted in Fig. \ref{fig:Schematic transfer}(a) as a function
of the energy difference between the states $E_{m_x}-E_{m_x'}$ for
$N=M=10,\Delta=0$. Transition amplitudes are large only between
dressed states of neighboring energies in the two subspaces. The
damping process is schematically shown in Fig. \ref{fig:Schematic
transfer}(b). Starting from the $N=M=10$ subspace with $11$
dressed states, Beliaev damping of an excitation from the
$\mathbf{k}$ momentum mode into two empty modes, $\mathbf{q'}$ and
$\mathbf{k}-\mathbf{q'} $, will transfer the system into the $N'
=9$, $M' =10$ subspace with  $10$ dressed states. Since the energy
spectrum in both subspaces is nearly linear, and there is one less
state in the smaller subspace, the energy differences between
states coupled by $H_{damping}$ are almost the same for all
initial dressed states. The decay of each dressed state into only
two neighboring states in Fig. \ref{fig:Schematic transfer}(b)
results in a structure of a doublet in the Beliaev damping
spectrum which is plotted in Fig. \ref{fig:damping spectrum} for
 experimentally realistic parameters.

We choose as a model system a condensate of $3\times10^{5}$,
$^{87}$Rb atoms in the $F=2$, $m_{f}=2$ ground state. The
condensate, which is similar to the experimental parameters of
\cite{ours-bragg}, is homogeneous and has a density of
$3\times10^{14}$ atoms/cm$^{3}$. The damping rate per excitation
of each transition is taken as a Lorentzian with a width and
height of $\Gamma/N_{\mathbf{k}}$ around $E_{m_x}-E_{m_x'}$.
$N_{\mathbf{k}}$ is the average occupation of mode $\mathbf{k}$
for the  dressed state $\ket{m_x}$. Averaging over all of the
possible transitions between subspaces, weighed by the initial
dressed state occupation, Fig. \ref{fig:damping spectrum} shows
the damping rate from the $N=M=5\times10^3$ to the
$N=5\times10^3-1$, $M=5\times10^3$ subspace vs. energy,  for
$k\xi=3.2$ , $q=k/\sqrt2$ and $\Delta=0$ (solid line). The rates
are calculated numerically according to the Fermi golden rule
\cite{ours-3wm}:
\begin{equation}
\Gamma_{m_x,m_x'}=\frac{2\pi}{\hbar}
\sum_{\vec{q'}}\abs{\bra{j-1/2,m_x'}H_{damping}\ket{j,m_x}}^2
\delta\left(\epsilon_k+E_{m_x}-\epsilon_q-\epsilon_{\vec{k-q}}-E_{m_x'}\right).\label{eq:decayrate}
\end{equation}
A clear doublet structure is evident. This splitting of the
spectrum is analogous to the Autler-Townes splitting where the
spectrum of spontaneous emission
 is split due to rapid Rabi oscillations \cite{Autler-Townes}.

Intuitively we expect a doublet in the spectrum to appear as a
result of the temporal oscillations in  the occupation of mode
$\vec{k}$. The decay from mode $\vec{k}$ is modulated, and the
spectrum of a linear system is a Fourier transform of it's time
correlation function \cite{tannor-book}. Temporal modulations,
therefore, lead to a splitting of the oscillation frequency  in
the decay spectrum. This argument is strictly true only for a
linear system such as $H_{2W}$.

 To understand the doublet structure of the decay spectrum, we use the local similarity of $H_{3W}$ to
$H_{2W}$, and approximate the three-wave damping rate by the
two-wave damping rate. The eigenstates of $H_{2W}$, $\ket{j,m_x}$
can be viewed as $N=2j$ spin $1/2$ quanta, $n_\alpha=j-m_x$ of
them pointing in the +x direction (right) and $n_\beta=j+m_x$ of
them pointing in the -x direction (left). We  use the Schwinger
boson mapping (sec. \ref{sec:schwinger}) to associate operators
$\ann{\alpha}{}$ and $\ann{\beta}{}$ with $\ann{J}{x}$. The state
$\ket{j,m_x}$ can be expressed by the quantum numbers
$\ket{n_{\alpha},n_{\beta}}$, just as the state $\ket{j,m_z}$ is
equivalent to $\ket{n_{\vec{k}},n_{\vec{k-q}}}$. The operator
$\ann{b}{\vec{k}}$ which annihilates a spin pointing in the +z
direction is then given by
$\ann{b}{\vec{k}}=1/\sqrt{2}(\ann{\alpha}{}-\ann{\beta}{})$. We
use the Fock representation to write
$\left(\ann{\alpha}{}-\ann{\beta}{}\right)\ket{n_{\alpha},n_{\beta}}
=\sqrt{n_{\alpha}}\ket{n_{\alpha}-1,n_{\beta}}-\sqrt{n_{\beta}}\ket{n_{\alpha},n_{\beta}-1}$.
Transforming this to angular momentum quantum numbers using
$n_{\alpha}=N/2+m_x=j+m_x$ and $n_{\beta}=j-m_x$, we get
\begin{equation}
 \ann{b}{\vec{k}}\ket{j,m_x}\approx\frac{1}{\sqrt{2}}\left(
 \sqrt{j+m_x}\ket{j-\frac{1}{2},m_x-\frac{1}{2}}-\sqrt{j-m_x}\ket{j-\frac{1}{2},m_x+\frac{1}{2}}\right).\label{eq:Hdecay}
 \end{equation}
 From Eq. (\ref{eq:Hdecay}) it is evident that
$H_{damping}$ couples the dressed state $\ket{j,m_x}$ only to the
two dressed states $\ket{j-1/2,m_x\pm 1/2}$.

 We have derived Eq.
(\ref{eq:Hdecay}) under the assumption that $H_{3W}$ is indeed
approximated by $H_{2W}$.  In Fig. \ref{fig:Schematic transfer}(a)
we see a comparison between the matrix elements of $H_{damping}$
for numerically obtained dressed states $\ket{j,m_x}$ (filled
circles) and the two-wave approximation (empty circles). The
spread in the energy difference is due to the non-linearity in the
spectrum of $H_{3W}$, however $H_{damping}$ is found to
significantly couple only dressed states with neighboring energies
as in the two-wave approximation.

\begin{figure}[tb]
\begin{center}
\includegraphics[width=8cm]{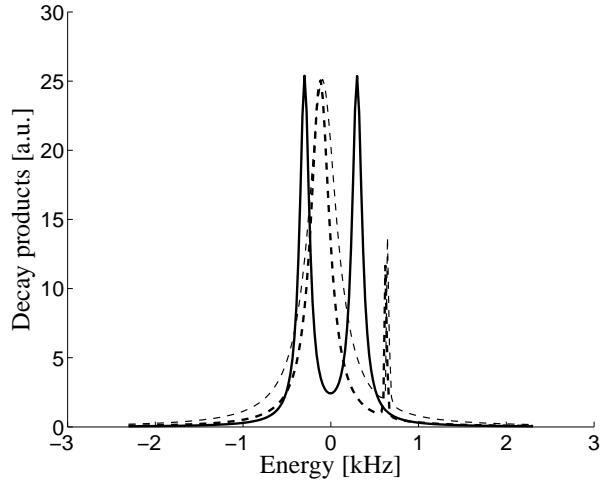}
\end{center}
\caption{Damping spectrum between the $N=M=5\times10^3$ subspace
and the $N=5\times10^3-1$, $M=5\times10^3$ subspace, at
$k\xi$=3.2, and $q=k/\sqrt{2}$. For these parameters
$\bar{\Omega}=2\pi\times 610$ Hz. For $\Delta=0$ (solid line), the
perturbative result  is indistinguishable from the numerically
calculated spectrum. For $\Delta=\bar{\Omega}$ the perturbative
result (thin dashed line) is still a reasonable approximation to
the numerical calculation (thick dashed line). The calculation was
done with $j=50$ and scaled to $j=2500$.} \label{fig:damping
spectrum}
\end{figure}

A perturbative result for the decay spectrum where energies are
calculated according to Eq. (\ref{eq:epert3w}) and rates according
to Eqs. (\ref{eq:decayrate},\ref{eq:Hdecay}) is indistinguishable
from the numerical result in Fig. \ref{fig:damping spectrum} for
 resonant wave mixing.

 Instead of one shell, which is the energy conserving surface in
momentum space for s-wave collisions, splitting in the Beliaev
Damping spectrum would direct the colliding atoms into two
separate shells \cite{ours-3wm}. Experimentally, the energy
doublet can be observed by computerized tomography analysis of
time of flight absorption images of the 3WM system
\cite{ours-tomography}.

In the presence of detuning, the Beliaev decay rate is calculated
in a similar manner, again using Eq. (\ref{eq:decayrate}).  A
damping spectrum for $\Delta=\bar{\Omega}$ is shown as a dashed
line in Fig. \ref{fig:damping spectrum}. The distance between the
peaks increases with the detuning to a value of
$\Omega_{eff}\approx\sqrt{\bar{\Omega}^2+\Delta^2}$.  As the
detuning is increased, one peak grows while the other shrinks. For
a positive (negative) detuning, the negative (positive) frequency
peak becomes larger, whereas the magnitude of the positive
(negative) energy peak decreases, for a large detuning, as
$1/\Delta^{2}$. The perturbative result (thin dashed line) still
fits the numerical prediction (thick dashed line) well. Deviations
of the perturbative result are due to the non-linearity in $m_x$
of the transition probabilities between three-wave dressed states,
while our model assumes these probabilities are linear as implied
by Eq. (\ref{eq:Hdecay}).

In Fig. \ref{fig:peak centers vs. detuning} we plot the spectrum
vs. the detuning, where energy and detuning are in units of
$\hbar\bar{\Omega}$. Darker areas in the graph correspond to
higher transition rates. For a positive (negative) detuning, the
center of the positive (negative) energy peak is approximately at
$E = \hbar\Delta$ (diagonal dashed line in Fig. \ref{fig:peak
centers vs. detuning}). The center frequency of the negative
(positive) frequency peak approaches $E=0$ (horizontal dashed line
in Fig. \ref{fig:peak centers vs. detuning}) like $1/\Delta$. For
a detuning larger than $\bar{\Omega}$, the center of the line is
well approximated by the perturbative result given in Eq.
(\ref{eq:AC Stark shift}).
\begin{figure}[tb]
\begin{center}
\includegraphics[width=8cm]{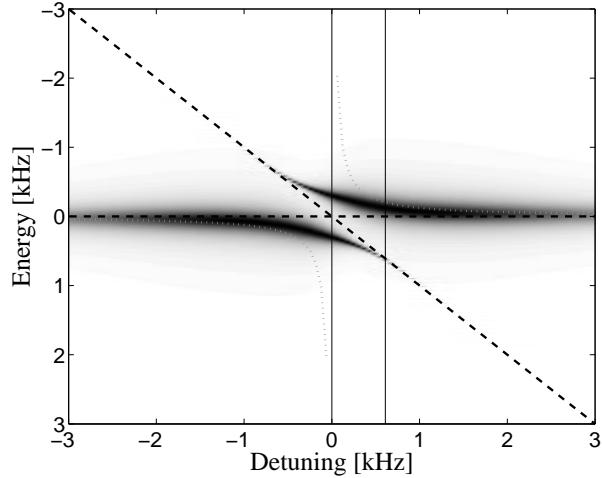}
\end{center}
\caption{The decay spectrum for different values of detuning.
Darker areas  correspond to higher transition rates. For large
positive and negative detuning, the spectra asymptotically
approach $E=0$ and $E=\hbar\Delta$ ( black dashed lines). The gray
dotted line is the perturbative result of Eq. (\ref{eq:AC Stark
shift}). The thin vertical lines are at the location of the
spectra in Fig. \ref{fig:damping spectrum}.} \label{fig:peak
centers vs. detuning}
\end{figure}

\begin{figure}[tb]
\begin{center}
\includegraphics[width=8cm]{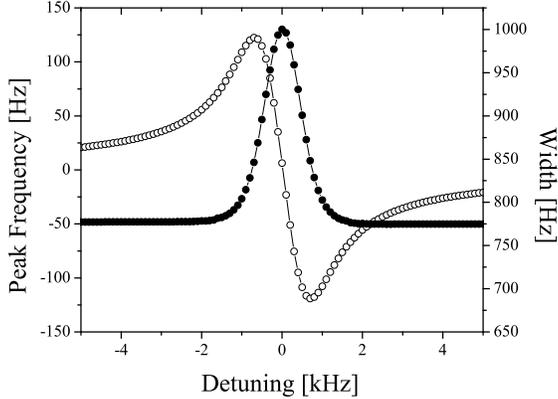}
\end{center}
\caption{The width (filled circles) and the peak location (empty
circles) of the spectrum plotted in Fig. \ref{fig:peak centers vs.
detuning}, in the presence of additional broadening taken to be a
gaussian of width $\bar{\Omega}$. Although the broadening washed
out the splitting, the broadening and resonance shift are still
apparent. \label{fig:broadening}}
\end{figure}

Several additional mechanisms will contribute to the broadening of
the two peaks beyond what is plotted in Fig. \ref{fig:damping
spectrum}. The fact that only the first scattering event occurs
between the $N=M=5\times10^3$ and the $N=5\times10^3-1$,
$M=5\times10^3$ subspaces will  broaden the resonances. Since the
energy splitting scales as $\sqrt{N}$, in an experiment where one
scatters $dN$ atoms from mode $\mathbf{k}$, this will result in a
relative broadening of ~$dN/2N$. According to the same scaling, an
initial coherent, rather than Fock, state will cause a relative
broadening of $\sqrt{N}/2N$. In the laboratory, the finite size
and the inhomogeneous density profile of the trapped condensate,
will cause additional broadening, and often dominate the  width of
the Beliaev damping spectrum. In this case the doublet structure
of the resonance may be unresolved, however the changes in the
transition energies and heights, are translated to changes in the
resonant frequency and the width of the collisional signal as
shown in Fig. \ref{fig:broadening}. There will be a broadening of
the decay spectrum  around $\Delta=0$, whereas the energy of the
collisional products will fall on  a dispersive curve around the
resonant value. Figure \ref{fig:broadening} indicates that even
though splitting and oscillations may be difficult to observe when
other decoherence mechanisms dominate over the 3WM coupling, line
shifts and broadening can be still easily observed under such
conditions.

Another way to measure this spectrum is by probing mode $\vec{k}$
with a weak Bragg pulse. A spectrum obtained by setting the
momentum of two laser beams, and sweeping the detuning between the
beams, around the resonance of a quasiparticle excitation of the
macroscopically occupied mode $\vec{k}$, is similar to that
plotted in Fig. \ref{fig:damping spectrum}. When the wave mixing
is resonant, there is a splitting in the spectrum, and when the
wave mixing is non-resonant, one peak dominates, and in effect the
spectrum is shifted as in the optical ac-Stark shift. We
calculated the Bragg spectrum for the case where the momentum mode
$\vec{q}$ is probed rather than $\vec{k}$, and found that the
spectrum would be a triplet rather than a doublet. This is in
analogy to the Mollow splitting in atomic physics
\cite{Mollow,API}.

In many experiments, the Beliaev damping is the main cause of
decoherence. From Eq. (\ref{eq:Hdamping}) it is clear that the
total damping rate decreases for small values of $k$. This is due
to a large reduction in the term $A_{\vec{k,q'}}$, and the smaller
number of available modes into which decay is possible $\vec{q'}$.
This suggests matter wave mixing experiments should be done in the
low momentum three-wave regime rather than the high momentum
four-wave regime.

 In this section we focused on the Beliaev damping of
three-wave mixing. The same arguments give rise to a splitting in
the energy of Beliaev damping from a rapidly oscillating BEC
discussed in sections \ref{sec:hamiltonians-2wm}, and
\ref{sec:2wm}. We studied the Beliaev damping of such a system
experimentally, and found a large deviation from the expected
s-wave sphere \cite{ours-collisions}.

\section{Conclusion}

In conclusion, we use the well known Schwinger boson mapping to
describe the spectrum and dynamics of wave mixing. Motivated by
matter wave mixing of Bose-Einstein condensates, we develop a
framework for describing two and three-wave mixing. We first
present the solution of a noninteracting two-wave mixing problem,
corresponding to Rabi oscillations between two discrete momentum
modes, induced by two-photon Bragg coupling. Interactions between
modes are then added and the exact solution, obtained by
diagonalizing the Hamiltonian is compared to a perturbative
expansion. When the Rabi oscillation frequency $\Omega$  is
smaller then the chemical potential of the condensate $\mu$,  the
interactions are well described as a perturbation over the
non-interacting solution, yielding a simple analytic expression
for the non-linear quantum dephasing time. This dephasing is due
to quantum fluctuations and cannot be described by mean field. The
spectrum of such an oscillating BEC, can be calculated using this
framework, and is found to agree with the measured one
\cite{ours-splitting}. For a larger chemical potential, the
perturbative solution breaks down, and differs from the exact
solution. For  $\hbar\Omega=\mu$  the GPE is dynamically unstable,
and cannot be used to predict dynamics. This case of two-mode
dynamical instability is different than that of
\cite{inguscio-instability}, which is an instability of a single
Bloch state, and is under current experimental investigation.

We then analyze the spectrum and dynamics of three-wave mixing of
Bogoliubov quasiparticles over a BEC. We study the spectrum and
dynamics by treating the three-wave mixing locally as a two-wave
mixing problem, multiplied by a factor representing the third
field. By comparison to direct diagonalization of the three-wave
mixing Hamiltonian,  we find this approximation to be valid even
for small seeds $M<N$. We derive analytic expressions for  the
spectrum and decay time of the wave mixing oscillations at
different detunings from the energy-conservation condition. We
consider the coupling to the empty modes, and describe the damping
from an oscillating mode due to collisions with the BEC. We use
the local similarity to two-wave mixing to explain the underlying
cause for the splitting in the spectrum of the Beliaev damping
products obtained numerically from the exact three-wave
diagonalization.


The Beliaev Damping phenomena studied in this paper, appear also
in two-wave mixing. A Bose-Einstein condensate undergoing rapid
Rabi oscillations, as presented in section
\ref{sec:hamiltonians-2wm},  exhibits splitting in the Bragg
spectrum of the different momentum modes, and in the Beliaev
damping as well. Both these phenomena have been recently observed
experimentally \cite{ours-splitting}.

 This dressed state
approach can be extended to four-wave mixing of momentum modes
$\vec{k_1},\vec{k_2}\vec{k_3}$ and $\vec{k_4}$, where
$\vec{k_1}+\vec{k_2}=\vec{k_3}+\vec{k_4}$ \cite{phillips-4wm}.
Here the conserved quantities are $N=n_1+n_2+n_3+n_4$, $n_1-n_2$,
and $n_3-n_4$. If initially two modes are macroscopically occupied
(say 1 and 3), the four-wave mixing is again similar to two-wave
mixing of modes 2 and 4.  Hence, one observes oscillatory dynamics
which lead to a splitting in the decay spectrum.

Another case of three-wave mixing of bosonic matter waves is
binary molecule formation. The Hamiltonian governing this process
has been studied analytically for the case where both atoms are in
the condensate mode \cite{vardi-bosonic_molecules}. Our model
corresponds precisely to molecules formed of two separate atomic
modes, and can be used to described the evolution of such a
system.

This work was supported in part by the Israel Ministry of Science,
the Israel Science Foundation and by Minerva foundation.

\end{document}